\documentstyle[12pt,aaspp4]{article}

\def\refit{{\null}}
\def\refbf{{\null}}
\def\etal{{\refit et~al.\ }}
\def\HST{{\sl HST\/}}
\def\IUE{{\sl IUE\/}}
\def\Halpha{{H$\alpha$\ }}
\def\Hbeta{{H$\beta$\ }}
\def\ie{{\refit i.e.,}\ }
\def\eg{{\refit e.g.,}\ }

\def\pmb#1{\setbox0=\hbox{#1}
  \kern-.02em\copy0\kern-\wd0
  \kern.01em\copy0\kern-\wd0
  \kern.01em\copy0\kern-\wd0
  \kern.01em\copy0\kern-\wd0
  \kern.01em\copy0\kern-\wd0
  \kern-.02em\raise.01em\box0 }

\begin{document}

\title{A \pmb{\sl Hubble Space Telescope\/} Survey for Resolved
Companions of Planetary-Nebula Nuclei}

\author{Robin Ciardullo\altaffilmark{1},
        Howard E. Bond\altaffilmark{2},
        Michael S. Sipior\altaffilmark{1},
        Laura K. Fullton\altaffilmark{2,3}, \\
        C.-Y. Zhang\altaffilmark{2,4},
        and Karen G. Schaefer\altaffilmark{2,5}}

\altaffiltext{1}
{Department of Astronomy and Astrophysics, Pennsylvania State University,
525 Davey Lab, University Park, PA 16802;
rbc@astro.psu.edu, sipior@astro.psu.edu}

\altaffiltext{2}
{Space Telescope Science Institute, 3700 San Martin Drive, 
Baltimore, MD 21218; bond@stsci.edu}

\altaffiltext{3}
{Current address:  Observatoire de Gen\`eve, CH-1290 Sauverny, 
Switzerland; Laura.Fullton@obs.unige.ch}

\altaffiltext{4}
{Current address:  Department of Physics \& Astronomy, University 
of Calgary, Calgary, Alberta, Canada T2N 1N4; zhangc@acs.ucalgary.ca}

\altaffiltext{5}
{Current address: Department of Physics, Astronomy, \& Geosciences, 
Towson University, Towson, MD 21252; kschaef@towson.edu}

\begin{abstract}

We report the results of a {\sl Hubble Space Telescope\footnote{Based on
observations with the NASA/ESA {\sl Hubble Space Telescope}, obtained at the
Space Telescope Science Institute, which is operated by AURA, under NASA
contract NAS 5-26555.}} ``snapshot'' survey aimed at finding resolved binary
companions of the central stars of Galactic planetary nebulae (PNe).  Using
the WF/PC and WFPC2, we searched the fields of 113 PNe for stars whose close
proximity to the central star suggests a physical association.  In all, we
find 10 binary nuclei that are very likely to be physically associated, and
another six that are possible binary associations.  By correcting for
interstellar extinction and placing the central stars' companions on the main
sequence (or, in one case, on the white-dwarf cooling curve), we derive
distances to the objects, and thereby significantly increase the number of PNe
with reliable distances. 

Comparison of our derived distances with those obtained from various
statistical methods shows that all of the latter have systematically
overestimated the distances, by factors ranging up to a factor of two or more.
We show that this error is most likely due to the fact that the properties of
our PNe with binary nuclei are systematically different from those of PNe
used heretofore to calibrate statistical methods.  Specifically, our PNe tend
to have lower surface brightnesses at the same physical radius than the
traditional calibration objects. This difference may arise from a selection
effect:  the PNe in our survey are typically nearby, old nebulae, whereas most
of the objects that calibrate statistical techniques are low-latitude,
high-surface-brightness, and more distant nebulae.  As a result, the
statistical methods that seem to work well with samples of distant PNe, \eg
those in the Galactic bulge or external galaxies, may not be applicable to
the more diverse population of local PNe. 

Our distance determinations could be improved with better knowledge of the
metallicities of the individual nebulae and central stars, measurements of
proper motions and radial velocities for additional candidate companions, and
deeper \HST\/ images of several of our new binary nuclei. 

\end{abstract}

\keywords{planetary nebulae: general --- binaries: visual --- stars: distances
--- stars: AGB and post-AGB}

\section{Introduction}

Planetary nebulae (PNe) are extraordinarily useful for probing stellar
evolution and cosmology.  In extragalactic astronomy, the planetary-nebula
luminosity function (PNLF) is one of the most accurate and reliable indicators
of relative distance (see the review of Jacoby \etal 1992); in stellar
astrophysics, PNe allow us to examine the physics of mass loss and the
timescales of stellar evolution (e.g., the review of Iben 1995).  Since young
PNe are bright emission-line sources, they make ideal test particles for
dynamical studies, both in the Milky Way and in external galaxies (\eg
Ciardullo, Jacoby, \& Dejonghe 1993; Amaral \etal 1996). Finally, since the
chemical abundances in a PN reflect the chemistry of the interstellar medium
at the time of its progenitor's formation (with some light species possibly
altered subsequently by stellar nucleosynthesis), these objects can yield
unique insights into galactic star-formation histories and chemical and
stellar evolution (cf.~Dopita \etal 1997). 

Almost every interesting quantity related to PNe in the Milky Way---their
space density, formation rate, Galactic distribution, sizes, ionized and
total nebular masses, contribution to chemical evolution, and the luminosities
and evolutionary states of their central stars---depends critically upon their
distances.  But, unfortunately, the distances to PNe within the Galaxy are
known only poorly. It is a remarkable irony that while the PNLF can be used to 
derive relative distances to external galaxies to better than $\sim 10\%$ 
(cf.~Jacoby \etal 1992), the distances to Milky Way PNe are typically known to 
no better than a factor of $\sim 2$ or worse (\eg Terzian 1993, 1997). In 
fact, of the $\sim 1100$ known Galactic PNe, only a dozen or so have distances 
that are reasonably well determined using direct methods.  For the rest, it is
necessary to use various statistical techniques. 

One fundamental, but rarely used, method for obtaining distances to Galactic
PNe is through the photometric parallaxes of resolved companion stars. 
Perhaps two-thirds of all stars are members of binary systems, and the
evidence suggests that the orbital period distribution of these binaries is a
Gaussian in $\log P$ centered at $P \approx 180$~years, with a dispersion of
2.3 in the logarithm of the period (Duquennoy \& Mayor 1991). Main-sequence
binaries with periods less than $\sim 1000$~days, but still wide enough for a
red giant to form, will eventually evolve through a common-envelope phase and
produce systems with dramatically shorter periods or even a coalesced binary
(see Bond \& Livio 1990; Yungelson, Tutukov, \& Livio 1993).  However, stars
with larger initial periods will not interact, and their separations will
actually increase with time, as a consequence of stellar mass loss.  As a
result, it is likely that nearly half of all planetary-nebula nuclei (PNNs)
have wide binary companions. 

Despite this expectation, only a few PNNs are actually known to have resolved
visual companions.  The best-known case is the nucleus of NGC~246, which has a
14th-mag K-dwarf companion $3\farcs 8$ away (Minkowski 1960; Cudworth 1973).
Fitting of this companion to the main sequence provides what is generally
accepted as one of the most accurate PN distances (quoted as 430~pc by Cahn,
Kaler, \& Stanghellini 1992, and revised to $495^{+145}_{-100}$~pc on the
basis of CCD photometry by Bond \& Ciardullo 1999), and one that is almost
always used as a primary calibrator of statistical distance methods.  Other
PNNs reported to have resolved companions include those of NGC 650-1, A~24,
A~30, and A~33 (Cudworth 1973), NGC~3132 (Kohoutek \& Laustsen 1977), NGC~6853
(Cudworth 1973, 1977), A~63 (Krzeminski 1976), K~1-14 (Kaler 1981), and PuWe~1
(Purgathofer \& Weinberger 1980), but at the time of this survey, only the
companion to NGC~3132 had photometry accurate enough to provide a reliable
spectroscopic parallax (Pottasch 1980, 1984). 

Here we present the results of a {\sl Hubble Space Telescope\/} (\HST\/)
snapshot survey of Galactic planetary nebulae, designed to detect and measure 
resolved binary PNN companions.  In \S 2, we describe the survey and
the photometric procedures used to measure all the stars present on our CCD 
frames.  In \S 3, we use these data to create a list of PNNs which have
visual companions deserving of follow-up observations.  In \S 4, we discuss
the issue of interstellar reddening, and compare measures of extinction
based on two-color photometry with estimates derived from the emission lines
of the nebulae.  In \S 5, we transform our \HST\/ stellar magnitudes to the 
standard Landolt (1983, 1992) system, and present $V$ and $I$ magnitudes for 
109 central stars.  Because of the superior resolving power of the \HST,
these magnitudes are generally better than PNN measurements made from the 
ground, especially for objects with $V > 15.4$.  In \S 6, we discuss our 
candidate binaries individually, and derive distances to those 
systems that are most likely to be physically associated. Finally, in \S 7, we
compare our distances to estimates based on statistical methods, and discuss
the implications our observations have for the Galactic PN distance scale. 

\section{Target Selection, Observations, and Reductions}

Our observations were carried out as a ``snapshot survey'' during Cycles~3
and~5 of the \HST\/ General Observer program. \HST\/ snapshots are short
exposures taken during occasional gaps that remain in the observing program
after as many primary scientific observations as possible have been scheduled.
The targets that are observed during these opportunities are selected at 
random from a list of candidates provided by the observers.

The PNe in our target pool were chosen using criteria designed to maximize our
chances of finding resolved companion stars.  First among these criteria was
that the objects in our survey had to be nearby. Our target selection was
heavily influenced by the list of very nearby PNe given by Terzian (1993), and
nearly all of our targets have statistical distances from Cahn, Kaler, \&
Stanghellini (1992; hereafter CKS) and/or Zhang (1995) that are less than
$\sim 3$~kpc. Almost as important was our Galactic-latitude criterion:  in
order to reduce contamination from superposed field stars, most of our targets
were chosen to have $|b| > 10^\circ$.   A third selection criterion was known
or suspected binarity of the central star. We included in our target list the
nine PNNs with reported visual companions (as listed above), along with A~34,
A~66, and Th~2-A, whose visual companions were noted by H.E.B. during
ground-based observations. In addition we included seven central stars that
have composite or late-type spectra but are unresolved from the ground, and
show no evidence for being extremely close binaries.  Finally, to increase the
usefulness of our dataset for comparison with other distance techniques, we
included six PNe with Very Large Array expansion distances (Hajian, Terzian,
\& Bignell 1993, 1995; Hajian \& Terzian 1996), 15 nebulae with distance
estimates based on model-atmosphere analyses of their central stars (M\'endez,
Kudritzki, \& Herrero 1992), and seven PNNs known to be very close binaries
(A~46, A~63, A~65, HFG~1, K~1-2, LoTr~5, and NGC~2346; cf.~Bond 1994), from
which distances can potentially be obtained from light-curve solutions. In
all, 144~PNNs were included in our input target lists. 

Our snapshot images were taken with the \HST\/ between 1993 and 1997.
The Cycle~3 (1993) observations were obtained with the Planetary Camera
of the original Wide Field and Planetary Camera (WF/PC){}.  These data 
consisted of single exposures through the F785LP (``$I$'') filter plus
occasional exposures through the F555W (``$V$'') filter, usually with the
PN central star positioned near the center of the PC6 CCD{}.  

The Cycle~5
(1995-97) data were obtained using the F814W (``$I$'') and F555W
(``$V$'') filters of the Wide Field Planetary Camera~2 (WFPC2), with the 
CCD gain set at 14 e$^-$/DN{}.  For these observations the PNNs were centered
on the Planetary Camera chip and usually imaged twice in each filter, with
the second exposure typically being $\sim 2$ times longer than the first.
The uneven exposure times optimized our ability to detect binaries.
We generally tried to scale our shorter integrations so that the PNN's image
fell just short of saturation (\ie $\sim 50,000$~electrons in the central 
pixel); this insured an accurate measurement of the PNN's magnitude and color, 
and enabled us to search for companions with $\Delta m \lesssim 4.5$ to
within $\rho\sim 0\farcs 05$ of the central star.  Conversely, we attempted to 
scale our longer exposures so that the central pixel of the PNN could receive 
as much charge as possible ($\sim 100,000$~electrons) without bleeding
significantly into its neighboring pixels.  By doing this, we could search
for companions almost $\sim 7$~mag fainter than the central star with 
$\rho\gtrsim 0\farcs 3$ separation.

Because of the limited time available per snapshot visit, we imposed a maximum
integration time of 900~s per filter.  The second, deeper exposures in each
filter described above were omitted if they would be longer than 500~s in
F814W, or 200~s in F555W{}. Exposures longer than 600~s were split into two
equal-length integrations to aid in cosmic-ray removal. Because of the
uncertainties of the ground-based PNN magnitudes used to calculate our
exposure times, we had anticipated that not all of our targets would be
optimally exposed. In fact, however, most of our program PNNs had useful
exposures, and we were able to perform a careful search for visual binaries
around almost every object. There was a guide-star acquisition error for our
observation of Lo~4, which resulted in our obtaining only a single F555W
exposure under gyro control. 

A summary of the WF/PC and WFPC2 observations appears in Tables~1 and 2.  Of
the 144 objects in the target list, 113 were actually observed: 26 with WF/PC,
84 with WFPC2, and 3 with both.  


Photometric reduction of our CCD images was accomplished using a combination 
of IRAF and DAOPHOT/ALLSTAR (Stetson 1987).  After the standard STScI pipeline 
de-biasing and flat-fielding, we removed charged-particle events (``cosmic
rays'') from the data.  For this task, the Cycle~3 and Cycle~5 frames 
were handled differently.  Because the Cycle~3 data consisted only of
single exposures through the F785LP and F555W filters, we could not 
simply ``stack'' these images and remove discrepant pixels.  Instead
we employed a semi-automatic iterative technique, which took advantage
of the fact that stars on the original WF/PC frames had aberrated 
point-spread-functions (PSFs).  First, we filtered our images using a 
$7 \times 7$ moving window, and masked all points that deviated by more
than $3 \, \sigma$ from the median defined by their surroundings.  We
then compared these filtered images to their parent images, and carefully
examined the cores of all stars on the frames.  If the filtering algorithm
had affected the cores, the original pixel values were manually inserted
back into the images.  In this way, the bulk of the cosmic rays were removed
without affecting the stellar photometry.

Reduction of the WFPC2 data proceeded in a more standard manner.  All the data
were first multiplied by a corrective image, in order to compensate for
the geometrical distortions present in the flat-field images.  We then used the 
DAOPHOT routine {\it find\/} to produce a list of candidate objects for each 
frame.  These lists were culled of cosmic-ray events and other spurious 
detections by first rejecting all objects that fell within the vignetted
areas of the frames, and then comparing the object lists derived from the
individual frames of a given target.  Objects found on the shorter exposures, 
but not on the longer ones, were discarded as cosmic rays.

Photometry of the WFPC2 data was performed via a PSF fitting technique.
Using all the frames of our survey, we identified $\sim 20$ bright, isolated
stars in each of the four CCD camera fields, and from these data we defined the 
instrument's position-dependent PSF{}. Once the PSF was defined, we used
it to derive preliminary instrumental magnitudes and create star-subtracted
images.  If any poor subtractions were found, due either to the presence
of an undetected close binary, or to a poorly determined nebular background,
the process was repeated with a modified star list and/or a model for the
nebular emission (cf.~Ciardullo \& Bond 1996).   The raw instrumental 
magnitudes derived by ALLSTAR were then scaled to an 11-pixel-radius aperture 
using an aperture correction derived from the bright stars, and corrected for 
the effects of finite charge-transfer efficiency following the prescription of 
Holtzman \etal (1995).  Finally, the WFPC2 magnitudes were corrected for a 
non-linear term related to integration time (Casertano 1997), and put on the 
photometric system defined by Holtzman {\refit et al.}  

Because the aberrated WF/PC images had a brighter limiting magnitude than
their WFPC2 counterparts, there were, in general, fewer stars on these frames
and saturation in the stellar cores was less of a problem.  Thus, our
photometric techniques were simpler. In most cases,
photometry was accomplished by summing the flux of each star within an
aperture of radius 5~pixels; only when the stellar separations were of the
order of a few pixels, or when the stellar cores were saturated, were
DAOPHOT's PSF fitting routines used. The small-aperture measurements were then
converted to total instrumental magnitudes using aperture corrections found
from bright stars, and scaled to the \HST\/ standard system using the
ground-based WF/PC calibration of Harris \etal (1991) and the flight
calibration of Hunter \etal (1992).  Finally, depending on the Julian Date of
the observation, an additional correction of up to $0.1$~mag was applied to
the F555W magnitudes, in order to account for the WF/PC chips' loss of
sensitivity with time due to contamination (cf.~Ritchie \& MacKenty 1993,
1994). 

\section{Detection of Binaries}

Because our data contain no information about proper motion or radial
velocity, our assignment of physical binarity has to rely solely on the
spatial coincidence of stars.  The probability of our identifying a binary PNN
depends critically both on the apparent separation of the companion star, and
on the field-star surface density.  (In other words, in a sufficiently crowded
field even a closely separated physical pair would be mistaken for an optical
double.)  To improve our chances of detecting visual binaries, our target
sample was weighted toward objects at high galactic latitude and toward nearby
objects, as described above. Still, because the surface density of bright
stars is much lower than for faint stars, our analysis is significantly more
sensitive to brighter companions than to fainter ones. 

To estimate whether a given apparent companion star located near a PNN is
physically associated with it, we calculate the Poissonian probability $P$
that a random field star as bright as, or brighter than, the companion would
be projected by chance within a radius $\rho$ of the PNN{}. Mathematically, 
\begin{equation}
P = 1 - \left(1 - {\pi \rho^2 \over A} \right)^N, 
\end{equation}
where $A$ is the total sky area surveyed by each WF/PC or WFPC2 frame, and $N$
is the number of stars within this area that are at least as bright as the
companion.  $P$ is a function of the local stellar surface density,
which we determined directly from star counts on the frames.  

If true physical binaries are common in our sample, there should be an excess
of pairs with very low values of $P${}.  (It was, of course, just such an
argument, first made by Michell 1767, that established the existence of
physical double stars.) In Figure 1 we plot the distribution of $P$ for our
sample of PNNs, using the stellar neighbor of each central star that has the
lowest value of $P$ (i.e., the highest probability of being physically
associated; generally, but not always, this is also the nearest neighbor of
the PNN){}. If all of our frames were populated entirely by randomly
distributed field stars, without any true physical pairs, the distribution
would be flat; instead, Figure~1 shows a large excess of companions with very
small chance probabilities, particularly for $P\le0.05$. This is a clear
demonstration that we are detecting true resolved physical doubles. 

We therefore chose all of the apparent binaries with a chance probability of
5\% or less for further investigation, and list them in Table~3.  Most of the
columns are self-explanatory. Columns~2 and 3 give the separation, $\rho$,
between the PNN and the candidate companion, and the J2000 position angle of
the companion with respect to the central star.  Column~8 presents the
probability $P$ of a chance projection, as calculated from equation~(1). To
give an indication of the uncertainty in the probability, we give in the final
column of Table~3 the value of $P$ calculated by increasing the local stellar
density by $1 \, \sigma$ above that deduced directly from the star counts. 

Table~3 is {\it not\/} necessarily a list of true binary PNNs: it is merely
the complete set of objects that, formally, have more than a 95\% probability
of being a physical pair. Since we imaged over a hundred PNNs in the course of
this survey, we can expect $\sim 6$ chance superpositions to be contained in
the list.  In \S 6 we will examine the properties of the PNNs in question and
attempt to select those systems that are most likely to be true physical
pairs. 

Returning to Figure~1, we note that in addition to the sharp peak at small
probabilities, there is a slight excess of objects with chance probabilities
between 0.05 and 0.40. This suggests that a few additional PNNs not included
in Table~3 also have physical companions. Unfortunately, because the ratio of
true binaries to chance superpositions in this probability range is low, the
only way to identify these objects would be through proper-motion and/or
radial-velocity measurements of the candidate companions. 

\section{Interstellar Extinction}

The first step in deriving distances from two-color photometry is to obtain
an estimate of the amount of interstellar extinction suffered by each object.
This can be done in two ways.  

The first method is to use the observed emission ratios (\ie the relative 
intensities of H$\alpha$, H$\beta$, and the radio continuum) in the PNNs'
surrounding nebulae.  The intrinsic \Halpha to \Hbeta ratio of a typical
$10,000$~K nebula is $\sim 2.86$ (\eg Brocklehurst 1971); by comparing this
value to the observed line ratio, it is possible to compute the logarithmic
extinction at H$\beta$, denoted $c$.  Similarly, an estimate of $c$ can be
obtained by comparing the nebula's radio continuum (measured at 5~GHz) to its
total emission at H$\beta$ (cf.~Milne \& Aller 1975).  Measures of $c$ exist
in the literature for a large number of objects, and a useful compilation
appears in CKS\null.  Unfortunately, because most PNe exhibit complex
multi-zone structures, the exact values to use for the intrinsic
H$\alpha$/\Hbeta and H$\beta$/radio ratios can be difficult to determine; in
addition, observational errors introduced by atmospheric dispersion,
background confusion (for radio-faint PNe), and uncertain emission-line
photometry (for large and low-surface-brightness PNe) can lead one astray.
Finally, as our {\sl HST\/} images of NGC~7027 and IC~4406 in Figure~2 show,
the dust within a PN can be extremely patchy. Thus adoption of a global
extinction value for the particular line of sight to the central star (or to
its companion) may not be appropriate. 

The second method of determining the extinction to a PNN is to compare the
observed color of the central star with that expected from a hot source.  
Simulations using the WFPC2 efficiency curves (Biretta \etal 1996) demonstrate
that a star whose optical continuum falls on the Rayleigh-Jeans tail of the 
Planck function should have an F555W$-$F814W color of $\sim -0.4$.  As the 
histogram of Figure~3 illustrates, this value is confirmed in our survey: the 
distribution of PNN colors observed with WFPC2 has a hard blue limit of 
${\rm F555W} - {\rm F814W} \sim -0.4$, and a long tail to the red caused by 
interstellar (and/or circumstellar) reddening.  Values for the extinction 
of each PNN can therefore be obtained directly from our measured colors. Of
course, some of the same uncertainties that affect the nebula-based extinction 
ratios apply here: spectroscopy demonstrates that there are real departures 
from Rayleigh-Jeans continua in some objects (\eg from Wolf-Rayet emission 
lines; see, for example, Smith \& Aller 1969; Heap 1982; M\'endez \etal 1986). 
Moreover, some PNNs have intrinsically red colors (either from an unresolved 
late-type companion, or because they themselves have evolved back to a 
``born-again'' red-giant phase).   Finally, even if all central stars were 
identical, small-scale non-uniformities in the dust distribution (such as seen 
in Figure~2) could cause 
a companion star to have a different reddening value than the hot component.

Figure~4 compares the nebular extinction measurements (taken from CKS and 
Kingsburgh \& Barlow 1994) with reddenings derived from our WFPC2 photometry 
using the assumption that the intrinsic ${\rm F555W} - {\rm F814W}$ color of 
the central star is $-0.40$.  
For the figure, $E$(F555W$-$F814W)
has been transformed to $E(B$$-$$V)$ using the extinction table for an
O6 star given by Holtzman \etal (1995).
In the figure, M~2-9 and NGC~2022 have not been 
plotted, since both of their nuclei are marginally non-stellar on our \HST\ 
frames, and are probably surrounded by thick circumstellar dust.  In addition,
PNNs with known composite or late-type spectra have been omitted
from the diagram.  In those cases where the PN has both a radio-flux and 
Balmer-decrement value for $c$, we have used the mean of the two 
measurements (except for Mz~2, whose Balmer-decrement extinction of zero is 
clearly unphysical given the object's other properties).

The line in Figure~4 plots the relation between $E(B-V)$ and $c$ derived
by reddening a 100,000~K Planck curve with the Cardelli, Clayton, \& Mathis
(1989) extinction law, folding the resultant energy distribution through the
F555W and F814W system response curves, and transforming from
$E(\rm F555W-F814W)$ to $E(B-V)$ using the table of Holtzman \etal (1995).
This relation has a slight curvature, due to shifts in effective filter
wavelength with increasing extinction (\eg Kaler \& Lutz 1985).  An excellent
approximation to this curve is 
\begin{equation}
E({\rm F555W-F814W}) = 0.913 \, c - 0.012 \, c^2.
\end{equation}


As can be seen in Figure~4, this law fits the data reasonably well in the 
mean, as the average difference between the two reddening estimators is 
$\Delta E(B$$-$$V) = 0.00 \pm 0.02$.   However, it is also obvious that
there is a substantial amount of scatter; the dispersion about the relation
is $\sigma_{E(B-V)} = 0.16$~mag.  Interestingly, this scatter is not confined 
to highly reddened PNe:  even for those objects with $c < 0.2$
the scatter is still substantial, $\sigma_{E(B-V)} = 0.13$.  

Without additional information, it is impossible to determine whether 
the scatter in Figure~4 is due principally to errors in the nebular
extinction values or uncertainties associated with our PNN colors.  In some
individual cases, we can test whether the nebular extinction is 
reasonable by computing the dereddened color of the central star:  if the 
application of $c$ results in a color that is significantly bluer than the 
Rayleigh-Jeans limit, then the extinction has obviously been overestimated. 
However, from the data at hand, there is no reason to believe that one 
extinction estimate is better than the other.  For simplicity, we therefore 
assume that both methods have similar uncertainties, of order $\sigma_{E(V-I)} 
\simeq 0.11$~mag.  If a planetary nebula has both a color-estimated reddening 
and a reasonable value for the nebular reddening, then we use the mean of 
these two numbers, and assign an uncertainty of $\sigma_{E(V-I)} =  0.11 / 
\sqrt{2} = 0.08$~mag.  If no measurements of the nebula exist, if the value
of $c$ leads to an implausible color for the central star, or if the
PNN is known to have a composite or late-type spectrum (or poor photometry),
then we use a reddening based solely on the remaining valid method, and assume 
$\sigma_{E(V-I)} = 0.11$~mag.

\section{Transformation to $V$ and $I$}

The final step before deriving PNN distances from binaries is to transform
the photometric systems defined by the WF/PC and WFPC2 filters to the
Johnson-Kron-Cousins $V$ and $I$ system as defined by the standard stars of
Landolt (1983, 1992).  For the WFPC2 data, this was done using the equations
given by Holtzman \etal (1995), which should be good to $\sim 2\%$ for 
colors in the range $-0.3 < V - I < 1.5$.   Note that the bluest of our PNNs 
actually lie slightly outside this range at $V$$-$$I \approx -0.4$, and the 
transformations for these stars are not formally applicable.  However, as
the Holtzman \etal data for the extremely blue stars of $\omega$~Cen and the
blue spectrophotometric standards Grw~$+70^\circ$5824, Feige~110, and
AGK~$+81^\circ$266 show, the equations relating F555W and F814W to $V$ and $I$
are well behaved at the blue end, and produce residuals that are no worse than
those for stars of moderate color.  Hence our mild extrapolation for the
central stars should be valid.

Transformation equations are also available for the original WF/PC filter
set (cf.~Harris \etal 1991; Saha \etal 1994).  However, these relations are
only defined for stars with $B$$-$$V$ colors between 0.0 and 1.6.  Since many
of our nuclei are much bluer than this, we chose not to use these equations
directly.  Instead, we re-derived the transformations for $V$ and $I$ by
combining the red-star data of Harris \etal (1991) with observations of
his six blue Landolt standards: G~162-66, GD~108, G~163-50, G~93-48,
Feige~67, and SA~114-750.  The resulting transformation equations, which
are valid in the range $-0.5 < {\rm F555W} - {\rm F785LP} < 3$ are
\begin{equation}
{\rm F785LP} - I = 0.054 - 0.154 \ (V - I) - 0.002 \ (V - I)^2 \label{eq1}
\end{equation}
\begin{equation}
{\rm F555W} - V = 0.001 + 0.061 \ (V - I) - 0.009 \ (V - I)^2. \label{eq2}
\end{equation}
Both equations have r.m.s.~residuals of 0.023~mag.

We note here that ground-based photometry of central stars within
high-surface-brightness nebulae can be extremely difficult, due to the
bright, irregularly distributed flux from their surrounding nebulae
(see Ciardullo \& Bond 1996 for a discussion of this well-known problem).  
However, the superior resolving power of the \HST\/ reduces this 
problem enormously.  Consequently, the $V$ and $I$ magnitudes
obtained in our survey represent a useful new database for future
PN analyses, even for the majority of stars that are not 
resolved binaries.

Our measured PNN $V$ magnitudes and $V$$-$$I$ colors 
are listed in Table~4; for reference, the nebular-based values of $c$
(which are taken from literature measurements of the Balmer-decrement and/or
radio flux density) are also tabulated.  Table~5 lists the PNN $I$ magnitudes 
for those objects not observed in $V$; except for the composite and late-type 
stars A~46, He 1-5, and HFG~1, these data were transformed
to the standard system using the nebular-derived value of $c$ and the
assumption that the intrinsic $V-I$ color of the objects is $-0.4$.
(Abell 46, He 1-5, HFG~1 were transformed using ground-based colors.)
In most cases, the magnitudes listed in the tables are accurate to better than
$\sim 0.05$~mag, and represent mean values derived from our two exposures.
In a few cases, partially saturated stellar images compromised our ability to 
derive an accurate magnitude for the PNN{}.  For the objects where this 
occurred,
we double-checked our measurements by deriving both 
a PSF magnitude and an aperture-photometry magnitude, and intercomparing
the results computed from our long-exposure frame with those obtained from the
short exposure.  Those PNNs with uncertain magnitudes and colors are
noted in Tables~4 and 5 with colons.

Two central stars, those of IC~4997 and NGC~5315, 
were so badly overexposed that we were 
unable to derive
$V$ and $I$ magnitudes; and two more, NGC~6309 and NGC~6369, were too badly 
overexposed in the $I$ band only.  For A~82 we are uncertain of the identity 
of the central star, as discussed in detail below.

Figure~5 compares our \HST\/ $V$ magnitudes to ground-based measurements
obtained by Kaler and collaborators (Shaw \& Kaler 
1985, 1989; Jacoby \& Kaler 1989) and by Tylenda \etal (1991).  The figure
demonstrates that for PNNs with magnitudes brighter than $V \sim 15.4$, the
agreement is generally good. There is no systematic error between
our measurements and those of Tylenda {\refit et al.,} as the mean difference
between the two magnitude systems is $0.00 \pm 0.06$~mag.  Moreover, although
our \HST\/ magnitudes are systematically fainter than the Kaler \etal
measurements, the offset is small: when the highly discrepant objects IC~3568, 
NGC~7662, and NGC~7008 (which we resolve as a binary with two comparably
bright components) are omitted, the mean difference between the two systems 
is $0.05 \pm 0.03$~mag.  We note that, in general, the scatter between our
measurements and those of Kaler \etal is significantly smaller ($\sigma \simeq
0.13$~mag) than that for Tylenda \etal ($\sigma \simeq 0.32$~mag).  But,
more importantly, both sets of residuals are larger than the internal errors
of the measurements would indicate.  Some of this additional scatter
undoubtedly comes from intrinsic variability in the PNNs themselves, as
Bond \& Ciardullo (1990) and others have shown that $\sim 0.1$~mag variability 
in PN central stars is not uncommon.  

For PNNs fainter than $V \simeq 15.4$, the agreement between the \HST\/ data
and the ground-based measurements is considerably poorer.  This is
presumably due to the increased importance of nebular contamination 
in the ground-based photometry.  The effect is particularly noticeable in the
Kaler \etal data, where $V$-band fluxes derived from aperture photometry
are overestimated by as much as $\sim 3$~mag. 

\section{Physical Associations and Optical Doubles}

The goal of this paper is to determine distances to PNe with resolved
binary nuclei by fitting the companion stars to the main sequence.
To do this, one should, in principle, use a main sequence appropriate for
the metallicity of each individual nebula.  However, of the candidates
listed in Table~3, fewer than half have nebular abundance measurements, and
of those, only $\sim 5$ have reliable data on atomic species that can 
reasonably be assumed to have been unaffected by nuclear processing.
We have therefore chosen not to attempt any metallicity correction.  Moreover,
since most PNe arise from an old-disk population, we have also
chosen not to use a main sequence based on observations of young, metal-rich
clusters such as the Hyades or Pleiades.  Instead for our distance estimates,
we use an $M_V$, $V$$-$$I$ main sequence that is derived from a spline-fit to 
a color-magnitude diagram of old-disk field stars.  This relation, which
has kindly been provided to us by H.~Harris (1998), is based on 
U.S. Naval Observatory (USNO) CCD parallaxes of faint field stars
(Monet \etal 1992; Dahn 1993; plus recent unpublished measurements),
USNO photographic parallaxes of bright stars (Dahn \etal 1988), 
and a combination of data on large-parallax stars from other observatories.
Since the dataset excludes known halo objects, high-velocity stars, and
binaries, the resultant main sequence should be applicable directly to our PN
companions.  The adopted USNO main sequence is given in Table~6.

As mentioned above, our candidate binary PNNs were selected purely on the
basis of spatial coincidence.  Hence it is likely that a few optical doubles 
are mixed in with the true physical pairs. Although it is impossible to
state with complete certainty whether any individual object is, or is not,
a binary, we can make some plausibility arguments based on the inferred
properties of the stars and nebulae.

Table~7 lists inferred properties of all of our candidate binaries, grouped 
according to our belief that the physical association is probable, possible, 
or doubtful. The reasons for our categorization are given case-by-case below.
To derive the distance moduli given in column~3, we dereddened the stars
using the adopted $E(B$$-$$V)$ of column~2 and the prescriptions given
by Harris \etal (1991) and Holtzman \etal (1995).  We then fit the dereddened
companion(s) to the USNO main sequence of Table~6.   Column~4 lists the
formal errors in the derived distance moduli; these were calculated by
combining in quadrature our photometric errors, the errors associated with 
the PNN reddening estimates, and the uncertainty in the main-sequence $M_V$ 
value.  Note that this last term, which we assume to be $\sim 0.3$~mag,
usually dominates the error.  Due to the spread of field-star metallicities 
and the effects of stellar evolution, the observed main sequence in the solar 
neighborhood has at least a spread of 0.3~mag (Perryman \etal 1995; Jaschek 
\& G\'omez 1998).  Consequently, although the mean main sequence at any color 
may be well defined, individual distance determinations which are based only 
on two-color photometry must have at least this amount of error.  In the future
it should be possible to refine our distance estimates via metallicities 
deduced from spectroscopy and/or Str\"omgren photometry of the companion 
stars, but for the present, our individual PN distance measurements carry this 
substantial uncertainty.

Columns~6 through 9 in Table~7 give various properties of the stars
and nebulae under the assumption the companion stars are, indeed,
physically associated with their PNNs.  Column~6 is the projected
PNN-companion star separation in astronomical units, column~7 is the
physical size of the PNN's nebula as derived from the angular diameters listed
in Acker \etal (1992), column~8 is the absolute $V$ magnitude of the PNN,
and column~9 is the nebula's absolute [\ion{O}{3}] $\lambda 5007$ magnitude,
based on the line fluxes given by Acker \etal (1992).  Following
Jacoby (1989), we define
\begin{equation}
M_{5007} = -2.5 \log F_{5007} - 13.74,
\end{equation}
where the monochromatic flux, $F_{5007}$, is given in ergs~cm$^{-2}$~s$^{-1}$.

A discussion of the individual objects appears below.

\subsection{Probable Physical Pairs}
{\it Abell 31:\/}  The central star of Abell~31 lies within an extremely
large (diameter $970\arcsec$) nebula, and has a very faint, close 
companion that is detected only on our $I$ frames.  The extremely 
red color of the companion, $(V-I)_0 \gtrsim 3.2$, places an interesting limit 
on the distance: in order to be on the main sequence, the companion star must 
be closer than $\sim 440$~pc.  This compares to statistical distance
estimates that range from 230~pc (CKS) to 1~kpc (Zhang 1995).  The small
projected separation derived for the pair ($< 115$~A.U.) and the very low 
probability for chance superposition (0.07\%) argue strongly for additional 
observations.

{\it Abell 33:\/}  The companion to Abell~33 was first detected by
Cudworth (1973); the separation and position angle of the pair has not
changed significantly since then.  The low probability of a chance
superposition and the reasonable implied parameters of the system
suggest that the stars are physically associated. Additional evidence for this
conclusion comes from the approximate agreement between our distance of
1.2~kpc and the statistical distance estimates of $\sim 0.7$~kpc (CKS), 
1.6~kpc (Maciel 1984), and 2.9~kpc (Zhang 1995).

We note here that Abell~33 (and Abell~24) have the largest reddening
discrepancies in our sample.  According to the observed $V$$-$$I$ color of 
the central star, the extinction towards Abell~33 should be close to zero.
However, Kaler, Shaw, \& Kwitter (1990) have used the nebular emission
lines to infer a total \Hbeta extinction of almost a magnitude.  Because
of the object's high galactic latitude, the low extinction values derived 
by earlier emission-line studies (Chopinet \& Lortet-Zuckermann 1976; Kaler 
1983), and the negative reddening implied by the ratio of radio to H$\beta$ 
flux (CKS), we have chosen to ignore the Balmer-line extinction measurement
for this object, and have used $E(V$$-$$I) = 0$ in our calculations.  

{\it K 1-14:\/} Prints of the Palomar Sky Survey (POSS) reveal that a pair
of stars separated by $9\farcs1$ is at the center of this nebula.
H.E.B. noted this fact $\sim 20$~years ago, as did Kaler 
(1981), and on this
basis we included the object in our program.  However, Kaler \& Feibelman
(1985) concluded that neither of these stars is the PNN: using both {\it 
International Ultraviolet Explorer\/} ({\it IUE\/}) observations and a 
large-scale ground-based plate provided to them by F.~Sabbadin, these authors 
concluded that the actual central star of K~1-14 is a fainter third object, 
lying {\it between\/} the two POSS stars.   Our \HST\/ images confirm that 
the Kaler-Feibelman conclusion is correct: an extremely blue object lies 
$\sim 2\farcs 4$ southwest of the brighter POSS star.  (The PNN is invisible 
on the POSS prints due to the overlapping images of the field stars.)  
Remarkably, the hot star itself has a very close, even fainter companion 
$0\farcs 36$ away.  The very small separation of {\it this\/} pair, coupled 
with the low stellar density in the field, argue strongly for a physical 
association.  Our main-sequence-fitting distance of $\sim 3$~kpc is in 
reasonable agreement with that determined from statistical methods (3.4~kpc, 
CKS; 5.3~kpc, Maciel 1984). 

{\it K 1-22:\/}  This system has a number of remarkable parallels to K~1-14. 
The discoverer of the PN (Kohoutek 1971) noted that three stars appear near
the center of the nebula, and proposed that the brightest of these was the
PNN{}.  Smith \& Gull (1975) confirmed that this star is very blue, but
Kaler \& Feibelman (1985) noted that the visual luminosity of the star
was too large to match the flux distribution extrapolated from the
\IUE\/ measurements.  Instead Kaler \& Feibelman suggested that a fainter
object $4''$ east of Kohoutek's candidate was the PN's true central star.
Our \HST\/ frames confirm that the original Kohoutek-Smith-Gull star is indeed 
the PNN, but its visual flux is augmented by a very close visual companion
$0\farcs 35$ away.  In fact, at $V=17.13$, the companion is so red 
($V-I=1.12$) that it is actually brighter than the PNN in the $I$ band. 
Given the low stellar density in the field, the pair almost certainly form
a bound system.  Our derived distance of 1.3~kpc is reasonably
consistent with distances based on statistical techniques (1.0~kpc, CKS;
3.4~kpc, Zhang 1995). 

{\it K 1-27:\/}  The faint companion to K~1-27 has the color of a late A~star;
consequently if the star is on the main sequence, its faint apparent magnitude
($V \simeq 21.3$) implies an implausibly large distance (55~kpc).  There is,
however, another possible solution for this system: the companion star could
be a white dwarf.  Instead of placing the secondary on the main sequence, we
obtained the distance given in Table~7 by putting the companion on the
white-dwarf cooling sequence defined by the hydrogen-rich white-dwarf models
of Bergeron, Wesemael, \& Beauchamp (1995).  The locus of points defined by
these models is in excellent agreement with that observed for field white
dwarfs (Monet \etal 1992), and, by adopting the curve, we derive a white-dwarf
absolute magnitude of $M_V = 12.77$.  If we use this value, and ignore the
nebular Balmer-line extinction estimate $c = 0.28$ (which leads to an
unphysically blue color for the central star), then we derive a distance to 
the system of $\sim 470$~pc. 

There is only one other distance estimate, statistical or otherwise, for
K~1-27.  By modeling the absorption lines from the hydrogen-deficient central 
star, Rauch, K\"oppen, \& Werner (1994) derived a distance to the object of
$1.29^{+1.05}_{-0.58}$~kpc.  This number, however, assumes $c = 0.28$; with
a more reasonable value of $c = 0.08$, their distance decreases by a factor
of $\sim 1.2$, and comes into marginal agreement with ours.  In most
other respects, our white-dwarf hypothesis is reasonable.  At an
absolute magnitude of $M_V = 12.77$, the cooling age of a $0.575 M_\odot$
companion white dwarf is $\sim 1$~Gyr (cf.~Bergeron, Wesemael, \&
Beauchamp 1995), and thus the star would not be in a particularly rapid
phase of evolution.  Similarly, if the distance is indeed 0.47~kpc, then
the derived values for the nebular size (0.1~pc) and binary
separation (260~A.U.) are also plausible.  The only potential problem lies in
the extremely small amount of flux radiated by the nebula in [\ion{O}{3}] 
$\lambda 5007$.  At an absolute [\ion{O}{3}] $\lambda 5007$ magnitude of $M_{5007} 
\sim 8.7$, the nebula would be a full 13.1~mag fainter than the bright end of 
the [\ion{O}{3}] $\lambda 5007$ planetary nebula luminosity function (Ciardullo 
\etal 1989; Jacoby \etal 1992), and the intrinsically faintest [\ion{O}{3}]
$\lambda 5007$ source in the Strasbourg-ESO planetary nebula catalog (Acker
1992).  Note, however, that the nebular and central-star properties of K~1-27
(Henize \& Fairall 1981; M\'endez, Kudritzki, \& Simon 1985) are very similar
to those of Abell~36, an extremely high-excitation object which, from its
[\ion{O}{3}] line flux and statistical distance, {\it is\/} the faintest [\ion{O}{3}]
$\lambda 5007$ source currently in the catalog ($M_{5007} \sim 7.2$).  In
fact, both Henize \& Fairall and M\'endez \etal have remarked that the
properties of the K~1-27 nebula and central star are more extreme than those
of Abell~36, and a large fraction of the nebula's oxygen is in O~IV{}. 
Consequently, the intrinsically small amount of [\ion{O}{3}] $\lambda 5007$
emission is entirely reasonable.  We consider this a likely white dwarf-PNN
binary system. 

{\it Mz~2:\/} This PN is 3 degrees from the Galactic plane, and only 31 degrees
from the Galactic center.  Thus, the field-star density in the region 
is extremely high: there are over 3400 stars recorded on our Wide Field
and Planetary Camera frames.  Nevertheless, only five of these stars are as 
bright as, or brighter than, our putative companion.  The relative scarcity of 
$I = 15.8$ stars, and the small ($0\farcs 28$) separation between the PNN and
the companion, leads to a very high probability (99.99\%) of a physical
association. 

Our distance to Mz~2 of $\sim 2.2$~kpc is in good agreement with most
statistical distance estimates (2.3~kpc, CKS; 2.4~kpc, Van de Steene \&
Zijlstra 1994; 2.7~kpc, Zhang 1995).  Kingsburgh \& English (1992) derive a
somewhat larger distance ($\sim 5$~kpc) from the PN's location on the
[\ion{O}{2}] density-ionized mass relation, but the subsequent classification
of the object as a Type~I PN (Perinotto \etal 1994) vitiates this analysis. 

{\it NGC 1535:\/}  The companion to this well-studied PNN has the colors
of an early G~star.  If the pair form a bound system, then the projected
separation between the two stars is $\sim 2400$~A.U., and the distance
to the binary is $\sim 2.3$~kpc.  This value is in excellent agreement not
only with most modern statistical distances (2.3~kpc, CKS; 2.0~kpc, Van de 
Steene \& Zijlstra 1994; 2.1~kpc, Zhang 1995), but with the distance derived
from the non-LTE model-atmosphere analysis of its central star (2.0~kpc,
M\'endez, Kudritzki, \& Herrero 1992).  This consistency, along with 
the small probability of a chance superposition, supports the interpretation 
that this is a true physical system.

{\it NGC 3132:\/}  The binary nature of this PNN was first discovered by
Kohoutek \& Laustsen (1977).  The companion star has an A0~V spectral type
(Lutz 1977), which implies that the PN was ejected from an even more massive
progenitor.  The hot star, which is only marginally resolved from the 
ground, is easily measured on our PC frame, and is 5.65~mag fainter in $V$ 
than its companion; if we place the A~star on the main sequence, we
obtain a distance to the system of $\sim 0.8$~kpc.  This is consistent with
estimates based on the interstellar reddening along the line of sight
(Gathier, Pottasch, \& Pel 1986) and the ground-based spectroscopic parallax 
of the A~star (Pottasch 1980). 

Unfortunately, there are two caveats that accompany our distance
measurement.  The first arises out of the definition of our field-star
main sequence.  For a PN to exist, the age of the system must be at least
$\sim 5 \times 10^7$~years (Bressan \etal 1993); this is a non-negligible
fraction of the main-sequence lifetime of the A~star.  Consequently, some
main-sequence evolution must have occurred, and the companion could be up
to $\sim 0.2$~mag brighter than its zero-age main sequence luminosity.  While
the effect is partially mitigated by our use of a field-star main sequence
rather than a zero-age main sequence, it is still likely that the companion
is older than a typical A-type field star.  Fortunately, the effect is small:
the difference between the expected mean magnitude of a group of A stars
of all ages, and that of a sample of stars with ages of at least
$\sim 5 \times 10^7$~years is only $\sim 0.05$~mag.

The second problem comes from the uncertainty in our reddening estimation.
On the lower main sequence, the $V$ vs.~$V$$-$$I$ reddening vector is roughly
parallel to the main sequence.  Consequently a small error in the extinction
makes little difference to a distance derivation: both $V_0$ and $M_V$
are modified in a similar fashion.  For A~stars, however, this is not the
case, as the slope of the main sequence is significantly steeper.  The 
result is that a small uncertainty in reddening translates into a large 
uncertainty in distance.  In the case of NGC~3132, the uncertainty in our
color-based reddening estimate, $\sigma_{E(V-I)} \sim 0.11$~mag, translates
into a $\sim 0.64$~mag uncertainty in distance modulus.  This dominates
the error given in Table~7.  Str\"omgren photometry of the A~star would
improve the distance estimate considerably.

{\it NGC 7008:\/}  It is somewhat surprising that the composite nature of
the nucleus of NGC~7008 was not detected prior to our survey.  Although the 
angular separation between the PNN and the companion is small ($0\farcs 4$),
the magnitude difference between the two stars is only $\sim 0.5$~mag in
$V$, and in $I$ the light from the companion star actually dominates.  By
placing the companion on the main sequence, we obtain a distance of
$\sim 0.4$~kpc, and an implied stellar separation of $\sim 160$~A.U.{}
This compares to a distance of $\sim 1.1$~kpc estimated from extinction
measurements along the line of sight (Pottasch 1983), and the CKS statistical
distance of $\sim 0.9$~kpc.  Because the probability of a chance superposition
of a $V \sim 14$~mag star near the PNN is extremely small, we consider this
a good binary-star candidate.

There is one caveat to our distance, however.  On both of our F814W frames,
the point-spread function for NGC 7008's companion star appears slightly 
broader than expected from a normal single star.  The effect is not large and
we cannot rule out the existence of an instrumental problem.  However, it is
possible that the companion is itself a binary, making the PNN a hierarchical
triple.  If this is the case, our distance to the object is an overestimate. 

{\it Sp 3:\/}  The companion in this system has the color of an F-type star
and is only $0\farcs 3$ from the PNN; when we place it on the main
sequence, we derive a distance of $\sim 2.4$~kpc, in good agreement with the
CKS statistical distance of 1.9~kpc.   This consistency, along with the small 
probability of superposition, indicates that this is a true physical 
association.

\subsection{Possible Physical Pairs}

{\it Abell 7:\/}  Our 200-s $I$-band Planetary Camera image shows only two
stars on the frame: the PNN and a very faint companion only $0\farcs 9$
away.  Since the stellar density in this field is low (it is at
$b = -30^\circ$), there is a very high probability that
the stars are physically associated.  Unfortunately, the companion is
$\sim 5.3$~mag fainter than the central star in $I$ and so red that
it was not detected on our $V$~frames.  This implies a $V$$-$$I$ 
color redder than 1.21 and an absolute magnitude fainter than $M_V = 6.7$.

From main-sequence fitting, we can derive only an upper limit to
Abell~7's distance of $\sim 13$~kpc.  This limit is not
particularly useful, as the size of the nebula ($760\arcsec$) and the
central star's white-dwarf spectrum (Liebert 1980) demand that the
object be quite nearby.  If the CKS statistical distance of $\sim 200$~pc is
accurate, then the projected separation between the PNN and the companion
star is only $\sim 200$~A.U{}.  Deeper images are desirable to investigate
this interesting object further. In the meantime we can only classify A~7 as a 
possible physical binary.

{\it Abell 30:\/} The companion to Abell~30 was found from the ground by
Cudworth (1973).  The position angle has not changed significantly over the 
ensuing $\sim 25$~years, and our main-sequence fitting distance ($\sim 2$~kpc) 
is consistent with the CKS statistical distance of 1.7~kpc.  On the
other hand, the relatively large separation ($5\farcs 25$) does allow
a $\sim2\%$ probability of a chance superposition.  Moreover, at a distance
of 2~kpc, the derived physical separation of the pair ($\sim 10,600$~A.U.) 
is starting to approach the observed 0.1~pc cutoff imposed by the Galactic 
tidal field (cf.~Bahcall \& Soneira 1981; Latham \etal 1984).  We therefore 
categorize A~30 only as a ``possible'' physical association.

{\it Abell~63:\/}  The central star of Abell~63 is the 11-hour eclipsing 
binary UU~Sge (Bond, Liller, \& Mannery 1978); the PNN's nearby companion
($2\farcs8$ away) was first noted during photoelectric observations by
Krzeminski (1976). In spite of the PN's low galactic latitude, $b=-3^\circ$,
we find only a $1.5\%$ probability that the resolved companion is a 
chance superposition.  Based on the nebular Balmer decrement, the extinction
to the object is $c=0.71 \pm 0.10$ (Walton, Walsh, \& Pottasch 1993); this
number is in excellent agreement with the value obtained by modeling the
2200~\AA\ interstellar
absorption feature (Walton \etal 1993; Pollacco \&
Bell 1993).  When we combine this extinction estimate with our \HST\
photometry, we derive a distance to the companion star of $\sim 1.2$~kpc.

From the run of stellar reddening versus distance in that part of the sky,
a foreground extinction of $c \sim 0.7$ ($E(B$$-$$V) = 0.5$) implies a
distance of $\sim 1$~kpc (cf.~Bond \etal 1978), in good agreement with
our findings.  However, analyses of the spectral type and color of the
back hemisphere of the extremely close eclipsing companion yield values that 
are two to three times larger ($\sim 3.6$~kpc, Walton \etal
1993; $3.2 \pm 0.6$~kpc, Pollacco \& Bell 1993; $2.4 \pm 0.4$~kpc,
Bell, Pollacco, \& Hilditch 1994).  These distances could be 
overestimates, if some of the extreme heating effects on the companion 
``leak'' around to the unilluminated back side.  Unfortunately, without
improved observations, we can only identify this as a possible physical
system.

{\it IC 4637:\/}  This new pair---which remarkably was never noted from the 
ground---was discovered on our original Cycle~3 F785LP WF/PC frame and
followed up with WFPC2 (F555W and F814W) observations.  The implied 
separation between the PNN and its companion, $\sim 1200$~A.U., is large, 
but not excluded by any means; similarly, the derived nebular size of
$\sim 0.05$~pc is small, but again not unreasonable.  Interestingly, the 
statistical distances to this object vary widely, from 0.8~kpc (Amnuel 
\etal 1984) to 2.3~kpc (CKS), but all estimates lie on the high side of 
our 0.5~kpc value.  Furthermore, the object's only individual distance 
determination, the non-LTE model-atmosphere value of M\'endez, Kudritzki, 
\& Herrero (1992), is larger still (3.3~kpc).  We are therefore led to
classify this object only as a possible physical binary.

{\it NGC 2392:\/}  Like Abell 7, NGC~2392 has a faint companion which is
barely detected on our $I$~frames, and is invisible in $V$.  Our
upper limit to the distance, $\sim 6.4$~kpc, is not particularly useful,
since both non-LTE model atmospheres (M\'endez \etal 1992)
and statistical techniques (CKS; Van de Steene \& Zijlstra 1994; Zhang 1995) 
place the object closer than $\sim 2$~kpc.  Deeper imaging could provide a
photometric distance, but for the present, the lack of a $V$ magnitude
forces us to categorize the object as a ``possible'' physical association.

{\it NGC 2610:\/}  Again, there is a close ($0\farcs 6$), extremely faint 
companion to this object that could not be detected on our $V$ images.
A much deeper image is needed to produce a meaningful distance estimate.
We again categorize this as a ``possible'' physical system.

\subsection{Doubtful Physical Pairs}

{\it Abell 24:\/}  The visual companion to this star was first noted
by Cudworth (1973).  We measure a separation of $3\farcs33$, in good 
agreement with the $3\farcs4$ reported by Cudworth.
However, our measurement for the position angle of the binary
differs by $5^\circ$ from his value; this
offset marginally exceeds the uncertainty in the old measurement
(Cudworth 1997).  Since for any plausible distance, this change in position
angle is larger than is possible from binary orbital motion, the difference
argues against the existence of a physical association.  Other facts
suggestive of a chance superposition include the extremely
large diameter derived for the nebula ($\sim 4$~pc) and the factor of
five difference between our putative distance ($\sim 2.4$~kpc) and distances
derived from statistical techniques (\eg 0.52~kpc, CKS).  Despite the
fact that the formal probability of a chance superposition is only $\sim 2\%$,
we believe that this object is an optical double.

{\it NGC 650-1:\/} The companion star of this object was first noticed by
Cudworth (1973).  However, our \HST\ images reveal that the companion
is itself a very close double, with a separation of only $0\farcs 16$.
Given the field star density of the region, the two stars of the companion 
almost certainly form a physical system, as the probability of a $0\farcs 16$
chance superposition is less than 0.1\%.  However, associating the 
companion binary with the PNN is more problematical.

The separation and position angle of the companion pair relative to the PNN 
is the same as it was 25 years ago; thus, despite the fact that the 
probability of a chance association is $\sim5\%$, 
this may argue for a physical
association.  However, if the system is a hierarchical triple, then it is a 
very peculiar one: although both components of the companion have the same 
color, $(V-I)_0 \approx 0.87$, one is 0.8~mag brighter than the other.  In
other words, both stars cannot be on the main sequence.  If the PNN is 
associated with this pair then two of the three stars of this non-interacting 
trinary happen to be in a phase of rapid evolution.

The hypothesis of a bound triple system runs into further problems when one
considers the age and metallicity of the stars.  NGC~650-1 is a metal-rich
Type~I planetary nebula, with a most likely progenitor mass $M \gtrsim 2
M_\odot$ (Peimbert \& Torres-Peimbert 1983).  Yet if we assume a solar-like
metallicity and place the fainter component of the close pair on the main
sequence, then the position of the brighter component in the color-magnitude
diagram demands that the system be impossibly old.  As the isochrones in
Figure~6 show, both of the close components could lie on or close to the same
isochrone if the $V-I$ colors were near the blue ends of their error bars, but
the age of the pair would still be far in excess of that of the PNN{}.  A
lower age for the close pair would be possible if a substantial amount of
additional reddening were adopted.  However, if that were the case, then 
the reddening would greatly exceed that of the PNN itself, and we would be
forced to conclude that the two systems were unrelated.  Another alternative
is that the fainter component of the close pair is on the main sequence, and 
the brighter component is itself an unresolved binary containing two 
equal-brightness main-sequence stars; although not impossible, this option 
would imply the unlikely conclusion that the close pair is actually a 
hierarchical triple containing three stars of essentially identical mass. 

The distance of 4.1~kpc given in Table~7 was derived by fitting the fainter
component of the close pair to the main sequence; it exceeds the statistical
distances (CKS: 0.7~kpc; Zhang 1995: 1.6~kpc; Van de Steene \& Zijlstra 1994:
1.3~kpc) by a substantial amount.  We therefore suspect that the very close
pair, whatever its astrophysical explanation, is probably significantly more
distant than---and older than---the PNN, and thus not physically
associated. 

{\it PuWe 1:\/}  A slightly brighter companion to the central star of this
object is visible on the Palomar Sky Survey at a separation of $5''$
(Purgathofer \& Weinberger 1980), but our WF/PC observations reveal that
the companion is itself resolved into two stars with a separation of 
$0\farcs 6$ arcsec.  Unlike the NGC~650-1 system, both cool components 
are apparently on the main sequence, as they produce a consistent set of 
distances.  Since the PN is an extremely large object that extends over
$10'$ on the sky (Purgathofer \& Weinberger 1980), we expect it to be
relatively nearby, and our derived distance of $\sim 280$~pc does not
contradict this hypothesis. 

Nevertheless, the PuWe~1 PNN is probably not associated with the close pair.
Based on the stellar density in the region, and the rather large 
($5\arcsec$) separation between the PNN and the companions, the likelihood of
a chance superposition in the region is relatively large, $\sim 3\%$. 
Moreover, PuWe~1 is one of the few PNNs with a reliable trigonometric parallax
measurement, $2.3 \pm 0.4$~mas (Harris \etal 1997). This is more than $2
\sigma$ smaller than we would predict based on our companion star
measurements.  Even more telling is that the proper motion of the close pair
appears to be significantly different from that of the PNN (Harris 1998). 
Thus, we believe the stars are not physically associated. 

\subsection{Noteworthy Optical Doubles and Unresolved Objects} 

Several other stars in our survey were included because they were identified
as candidate visual binaries on the basis of ground-based observations, but we
have concluded that they are not physical pairs.  In addition, a few other
objects are definitely binaries, based on their composite spectra and/or red
colors, but we did not resolve them.  These objects are discussed in this
subsection. 

{\it Abell~78:\/}  
In his survey of the nebular features of this PN, Jacoby (1979) noted that
an extended patch of H$\alpha$ emission $\sim 10\farcs 2$ from the
PNN was spatially coincident with a star, but that there was no other
evidence for a physical association.  Our star counts in the region 
suggest that the ``companion'' is most likely an unrelated field star which 
is merely projected onto the nebula.

{\it Abell 82:\/} A~82 has a diameter of about $90''$, and was included in our
program as a candidate binary because, as recounted by Kaler \& Feibelman
(1985), there is a relatively bright late-type star at the center of the
nebula.  Kaler \& Feibelman's \IUE\/ spectra did reveal a weak, apparently
reddened, UV continuum within the $10''\times 20''$ aperture of the
instrument, but the derived color temperature was much too low to explain the
nebula's ionization. The central object has a K-type spectrum (Kaler \&
Feibelman 1985) and colors of $V=14.90$, $B$$-$$V=1.28$, $V$$-$$I=1.36$
(Kwitter, Jacoby, \& Lydon 1988, and our \HST\/ measurements). Kwitter \etal
have suggested that the image of the K~star masks the true PNN on ground-based
photographs (cf.~the case of K~1-14 above), but we find no evidence on our
\HST\/ frames for any blue object near the center of the nebula. 
Alternatively, Kaler \& Feibelman (1985) have proposed that a faint object
$6''$ northwest of the K~star is the PNN, but our photometry ($V=18.15$,
$V-I=1.10$) demonstrates that this star is not blue either.  In fact, the
bluest object on our PC frame is a $V=12.85$, $V$$-$$I=0.35$ star located
$18''$ southeast of the K~star. This star might have been within the \IUE\/
aperture if the pointing was slightly inaccurate, and thus could be the
reddened source detected by Kaler \& Feibelman. However, it is too far
off-center to be the planetary's central star. 

In view of the above ambiguities, we investigated the \IUE\/ observations in
further detail, with the aid of copies of the original observing scripts which
were kindly provided by W. A. Feibelman. Two short-wavelength spectra were
obtained, SWP~19771 (1983 Apr 20, 25~min), which shows no convincing
detection, and SWP~19908 (1983 May 5, 120~min), which detected the reddened
continuum described above, and shows that the source was well-centered in the 
aperture. For both observations, a blind offset was done from a nearby
11th-mag star onto the coordinates of the center of the PN{}.  Our
measurements on the Digitized Sky Survey show that the resulting pointing
would have been about $10''$ directly west of the 14th-mag K-type star, and
thus would either have missed it, or at best have had it at the edge of the
aperture.  However, handwritten notes from the telescope operator indicate
that a possible additional telescope movement may have been performed onto a
``second central star,'' which we speculate could have been the 12.8-mag star
to the southeast of the K~star.  In this case, the weak UV spectrum would be
nicely explained. If so, however, it appears that \IUE\/ never actually
observed the star at the center of the PN{}. 

We are thus left with two plausible conclusions: either the true hot central
star is a faint, unresolved companion of the K-type star; or a ``born-again''
scenario (see next paragraph) for the K~star itself may have to be invoked.
Further ground- and space-based spectroscopic observations of this PNN are
urged. 

{\it He~1-5\/ \rm and \it H~3-75:\/} The central star of He~1-5 is the 
well-known object FG~Sge, which appears to be a PNN that has undergone a 
late helium thermal pulse and has become a ``born-again'' red giant.  In 
agreement with spectroscopic observers (\eg Feibelman \& Bruhweiler 1990), 
and in agreement with the born-again scenario, we find no evidence for a hot 
companion in our \HST\/ images.  H~3-75 is an interesting and possibly related 
object:  according to Sanduleak (1984) the central star has a K-type spectrum.
{\it IUE\/} ultraviolet spectra (Bond 1993) show no trace of a hot
star, and our \HST\/ frames likewise show no resolved companion.  The
PNN is therefore a strong candidate for another born-again giant.
Spectroscopic observations are needed.

{\it He 2-36:\/} The optical central star of He 2-36 has a spectral type of 
A2~III (M\'endez 1978), and {\it IUE\/} observations suggest the presence of a 
hot companion (Feibelman 1985).  Our frames do not resolve the binary.

{\it M 1-2:\/} This object has a G2~Ib spectral type (O'Dell 
1966), and strong forbidden and permitted emission lines reminiscent of those
seen in a planetary nebula (see Grauer \& Bond 1981 and references therein). 
Our \HST\/ observations do not resolve any binary companion, nor do they show 
a resolved nebular component.  The system is probably a symbiotic-like binary
which is too compact to be resolved by \HST, although Feibelman (1983) has
argued from {\it IUE\/} observations that the star is surrounded by a young
planetary nebula.

{\it NGC 1514:\/}  The central star is a well-known composite system, 
containing a hot sdO star and an A-type companion (Kohoutek \& Hekela 1967; 
Greenstein 1972).  Greenstein's radial velocity measurements indicate that 
the period of the system must be quite long (or perhaps that the binary is 
seen nearly pole-on). Nevertheless, our \HST\/ images fail to resolve the 
system, setting an upper limit of approximately 40 A.U. for the projected 
separation. 

{\it NGC 6853:\/}  The binarity of the PNN was first suggested by Cudworth 
(1973), who identified a $V \simeq 17$ companion $6\farcs 5$ from the central 
star.  Astrometric measurements (Cudworth 1977) support this contention, as 
the proper motion of the companion is similar to that of the central star. 
Unfortunately, given the density of field stars in the region, our probability 
calculations cannot confirm this claim of binarity, as there is almost a 90\% 
chance that a random field star will be projected within $6\farcs 5$ of the 
PNN{}.  In fact, based on the star counts, the best candidate for association 
with NGC~6853 is a $V \approx 18.7$ star $1\farcs 1$ from the central star,
but even this object has a 12\% probability of being a chance superposition.  
We do note that if we assume Cudworth's star is associated with the PNN, then 
our \HST\/ photometric values of $V = 16.91$, $V$$-$$I = 1.83$, coupled with 
the assumption of no foreground reddening, leads to a distance of 
$430 \pm 62$~pc.  This is in agreement with the distance of $380 \pm 64$~pc 
recently obtain by Harris \etal (1997) from USNO parallax measurements.  In 
keeping with the precepts of this paper, we will not discuss this object any 
further, but we urge radial-velocity measurements for the system.

{\it Th 2-A:\/} As noted in \S 2, this object was included in our program
because of a nearby companion noted on ground-based CCD frames.  However, 
there are over 2300 field stars present on our \HST\ frames, and the star
in question has a $\sim 50\%$ probability of being a chance superposition.  
We therefore cannot classify it as a possible visual binary.

{\it A~34 and A~66:\/} Like Th 2-A, nearby companions were noted during
ground-based observations.  The companion of A~34 has a 28\%
probability of being there by chance, and for A~66 the probability is 21\%. 

{\it A~46, A~65, HFG~1, K~1-2, LoTr~5, \rm and \it NGC~2346:\/} All of these 
central stars (along with A~63, discussed above) are known to be extremely 
close binaries, based on their photometric variability (cf.~Bond \& Livio 
1990). Not unexpectedly, none of them were resolved in our survey, and we do 
not find any nearby resolved companions that have a high probability of being 
physically associated.

\section{Comparison with Statistical Distance Scales}

Due to their complexity and vast range in size, luminosity, mass, and
excitation, there is no reliable method for obtaining distances to large
samples of individual Galactic planetary nebulae.   As a result, in order to 
investigate planetary nebulae as a class, it is necessary to rely on
statistical distance estimators.  The principle behind these statistical 
distances is straightforward: the emitted Balmer-line flux from an ionized 
plasma depends almost exclusively on the total mass of the emitting region and 
the plasma density.  Consequently, if the ionized mass of a nebula can be
estimated, then its observed flux and angular size can be used to calculate
its distance. The key, of course, is to know the amount of ionized mass
contained in the nebula. 

There are several prescriptions in the literature for estimating this
mass, starting with the original assumption by Shklovsky (1956) that the
ionized masses of all PNe are the same, and that all PNe are optically thin.
Other formulations include adopting an ionized mass that is (a)~linearly 
proportional to nebular radius (Maciel \& Pottasch 1980), (b)~proportional to 
a power of the radius (Zhang 1995), (c)~proportional to the radio brightness 
temperature of the nebula (Van de Steene \& Zijlstra 1995; hereafter VdSZ), 
(d)~dependent on the [\ion{O}{2}]-derived nebular density (Kingsburgh \&
Barlow 1992; Kingsburgh \& English 1992), or (e)~constant for optically thin
nebulae, but proportional to an optical-thickness parameter for denser objects
(Daub 1982; CKS){}.  Once calibrated, each of these relations is capable of
producing distance estimates to large numbers of objects. 

Unfortunately, the number of PNe with independently known distances, which can
therefore be used as zero-point calibrators for these methods, is extremely
small.  Moreover, some of the calibrators
have distances that are themselves controversial.   For example, CKS
used 19~PNe with ``well-determined'' distances to calibrate their distance
scale, while VdSZ used 23 calibrators. A comparison of the two samples, 
however, reveals that only 16~PNe are common to both datasets, and of those, 
two objects have adopted distances that differ by more than a factor of two!  
A major reason for this dichotomy is that many of the ``well-determined'' PN 
distances are based on such methods as reddening of field stars projected 
near the PN line of sight (Gathier, Pottasch, \& Pel 1986), Galactic 
\ion{H}{1}
absorption measurements (Gathier, Pottasch, \& Goss 1986), nebular expansion 
parallaxes (\eg Hajian \etal 1995), and non-LTE atmospheric 
modeling of PN central stars (\eg M\'endez \etal 1992).  None 
of these methods is unassailable, and each carries its own (possibly 
substantial) uncertainty.

Our new 
sample of PNe with visual-binary nuclei significantly increases the number
of objects with reliable distance measurements, and constitutes a new
and important set of data with which to calibrate PN statistical distances. 
In addition to having quantifiable errors, the PNe in our sample have
distinctly different selection criteria from those measured by other methods.
PNe with interstellar-medium-based distances are mostly distant objects
in the plane of the Milky Way.  Similarly, PNe with nebular-expansion
distances are objects that are bright and optically thick, while those
analyzed with non-LTE model atmospheres have highly evolved central
stars.  Our wide binary stars, however, are primarily nearby objects and
objects at high galactic latitude.  Consequently, our sample not only enlarges
the PN calibrator database, but also reduces the possibility of a systematic
error due to selection biases.

The usefulness of our dataset for testing statistical distance techniques is
demonstrated in Figure~7, which compares directly measured PN distances with
those from four different statistical methods. For the directly measured
distances, we use the 10 ``probable'' associations listed in Table~7 (one of
which, A~31, is only an upper limit), along with the three additional
distances to the ``possible'' associations which are not upper limits. To
these we add the new ground-based distance to NGC~246 (based on photometry of
its wide binary companion; Bond \& Ciardullo 1999), and trigonometric
distances to seven PNe derived from recent $> 3 \sigma$ parallaxes measured by
the {\it Hipparcos\/} satellite (ESA 1997) and the U.S. Naval Observatory
(Harris \etal 1997). 

These accurate photometric and geometrical distances are plotted against
statistical distances computed from the 5~GHz flux measurements and
angular diameters given by Zhang \& Kwok (1993) and CKS, using the 
prescriptions of CKS, VdSZ, Maciel \& Pottasch (1980), and Zhang 
(1995).  For reference, the data of Figure~7 are given in Table~8.

It is immediately obvious that distances from all of the statistical methods
have considerable dispersion. Compared to the binary and astrometric
distances, the CKS and Zhang estimates scatter by $\sigma \sim 1.7$~mag and
$\sim 1.8$~mag in distance modulus, respectively.  The VdSZ distances exhibit
the smallest dispersion, $\sim 1.6$~mag, while the Maciel estimates have the
largest, $\sim 2.4$~mag.  (This last result is not very surprising, since some
of the PNe considered here have radii outside the formal limits of the Maciel
calibration.) Interestingly, a significant amount of dispersion is
attributable to one object, PHL~932, which has a {\it Hipparcos\/} parallax
distance of $110^{+48}_{-26}$~pc, but statistical distances that range from
800~pc (CKS) to 5.0~kpc (Zhang).  The central star of PHL~932 is an
exceptionally unusual object.  With a sdB spectral type (M\'endez \etal 1988),
it is one of only two known PNNs of this class, lying well off the normal
post-AGB evolutionary tracks.  It has been suggested that the star may have
evolved through a common-envelope binary interaction (Iben \& Tutukov 1993),
and thus the ionized mass could differ substantially from that of normal PNe.
If PHL~932 is arbitrarily disregarded, then the dispersion in the VdSZ and
Zhang errors drops dramatically to $\sim 1.1$~mag and $\sim 1.3$~mag,
respectively.   Even this, however, is much larger than the errors expected
from the techniques. 

Even more surprising are the zero-point offsets in the scales exhibited in the
figure.  All four of the statistical methods examined here systematically
overestimate the distances to the objects in our sample.   CKS come closest to
reproducing our distance scale, with estimates that are, on average, only
$\sim 25\%$ larger than the photometric and geometrical measurements.  Since
this is a $1 \, \sigma$ result, their analysis is still consistent with our
measurements.  The distance scales defined by VdSZ, Maciel, and Zhang,
however, are all significantly too long, with mean distance moduli that are
1.6, 1.2, and 2.7~mag larger than our own.  The sizes of these offsets are
extraordinary, especially when one considers that the VdSZ and Zhang relations
work well for samples of PNe in the Galactic bulge. 

Another way of looking at the problem is to compare the properties of our PNe
with those of other PNe with ``well-determined'' distances.  Following VdSZ,
we plot in Figure~8 the distance-independent PN radio brightness temperature
(which is related to the CKS optical depth parameter), against the
distance-dependent quantity of PN radius.   The filled circles represent
planetary nebulae with resolved binary companions, the crosses show PNe with
trigonometric-parallax distances, and the open circles are the VdSZ sample of
PNe with what they considered to be ``well-determined'' distances. It is clear
from the figure that the PNe in our new sample do not obey the rather tight
relation defined by the VdSZ calibrators; at a given radius, the binary and
astrometric PNe are systematically fainter in the radio and have a larger
amount of scatter (although the scatter is dominated by a few outliers). 

We interpret Figure~8 as revealing a classical selection effect. The distances
for the PNe in the VdSZ sample come almost exclusively from reddening and
\ion{H}{1} absorption distance determinations; hence these objects are nearly
all relatively bright, high-surface-brightness PNe that can be seen at great
distances along the plane of the Milky Way.  Consequently, a calibration using
this sample of objects nicely recovers the distances to Galactic bulge PNe.
The planetaries in our sample, however, are primarily nearby, faint, and
optically thin.   They represent a population of objects that has apparently
received little weight in the calibration of statistical distances. 

Figure 8 also points out the probable cause of a long-standing controversy
about the Galactic PN distance scale.  For years, there have been two distance
scales for Milky Way planetaries.  The traditional ``short'' distance scale
adopted by Cahn \& Kaler (1971) is supported by the extinction-distance
relation of Pottasch (1984), the Magellanic Cloud observations of CKS and
Webster (1969), and the Galactic bulge measurements of Stasi\'nska \etal
(1991).  All of these methods use bright PNe, similar to those analyzed in the
VdSZ study.  The opposing ``long'' distance scale, which is larger by a factor
of $\sim 1.5$, is supported by statistical-parallax measurements (Cudworth
1974), stellar-atmosphere models (M\'endez \etal 1992), [\ion{O}{2}]
line-ratio density estimates (Kingsburgh \& Barlow 1992; Kingsburgh \& English
1992), and number counts of PNe in other galaxies (Peimbert 1990).  These
techniques study a different sample of PNe, and include local,
lower-surface-brightness objects. Based on the results of Figures~7 and 8, it
is therefore not surprising that a different distance scale is derived. 

It is unfortunately less obvious---and beyond the scope of this paper---how
one could devise a new ``grand-unification'' calibration that simultaneously
handles both the lower-surface-brightness objects that prevail among the
nearby nebulae, and the brighter PNe that dominate samples like those in the
Galactic bulge and extragalactic systems. We leave this daunting task to 
future workers.

\section{Conclusion}

We have successfully used a large-scale \HST\/ snapshot survey to find 19
resolved companions of central stars in planetary nebulae. We consider ten of
these systems to be probable physical associations, another six to be possible
associations, and the remaining three to be doubtful. 

By fitting the companions to the main sequence (or in one case the white-dwarf
cooling sequence), we have derived reliable distances to the PNe.  Comparison
with various statistical distance estimates reveals that all of the current
statistical methods {\it overestimate\/} the distances to our sample. A more
detailed examination suggests that the well-studied nebulae used as
calibrators for statistical methods are biased toward high-surface-brightness,
low-Galactic-latitude objects. Our sample, on the other hand, contains more
objects of lower surface brightness, which may have systematically lower
nebular masses.  It will be a challenge to future refinements of the
statistical methods to include an additional correction for this effect. 

The primary source of error in our PN distance measurements from resolved
binaries is the unknown metallicities of the companion stars.  This
uncertainty propagates directly into the definition of the $M_V$, $V$$-$$I$
main sequence used to obtain the absolute magnitudes (and distance moduli) of
the stars.  Our distances could therefore be improved substantially via
abundance analyses of the nebulae and/or stellar atmosphere analyses (or
intermediate-band photometry) of the companion stars.  In addition, since
several candidate companion stars were detected only in $I$, deeper multicolor
imaging with \HST\/ (possibly with the restored NICMOS camera) could add
significantly to the list of PNe with direct distance determinations. Finally,
in order to confirm our candidate companion stars and identify additional
ones, proper-motion and radial-velocity measurements are needed. 

\acknowledgments

We thank H.~C.~Harris for providing a definition of the field-star $M_V$,
$V$$-$$I$ main sequence and for unpublished information. We would also like to
thank G.~Jacoby for instructive conversations at the start of this project and
on the properties of optically thin and thick nebulae, and for his
participation as a Co-Investigator on the Cycle~3 portion of the project.
S.~Torres-Peimbert provided support at a critical moment. This work was
supported by STScI grants GO-04308.01-92A and GO-06119.02-94A, and NSF NYI
grant AST 92-577833.

\newpage

\figcaption[figure1.ps] 
{Histogram showing the distribution of probabilities for the hypothesis
that the most-likely companion to each of our PNNs is actually a randomly
superposed field star.  If physical binaries did not exist, the distribution
would be flat.  Instead, there is a large excess of PNNs with nearby
companions ($P \le 0.05$).  These objects are listed in Table~3 and are
discussed in this paper.  The small excess in the range $0.05 < P < 0.40$
suggests that some additional physical pairs with larger separations
still await discovery.}

\figcaption[figure2.ps]
{\HST\/ WFPC2 $V$-band images of NGC~7027 (top) and IC~4406 (bottom).  For
NGC~7027 a logarithmic stretch has been used to display the image, due to the
large range of surface brightness.  Angular widths of the images are $23''$
(top) and $33''$ (bottom). Note, at \HST\/ resolution, the presence of
extremely patchy and filamentary dust features in both planetary nebulae.
Because of this small-scale structure, an extinction measurement based upon
global nebular properties, or even upon the colors of a central star, might
not be appropriate for a nearby resolved companion.} 

\figcaption[figure3.ps]
{The distribution of colors for PNNs observed with WFPC2, excluding those
objects with composite or late-type spectra.  The blue limit of the 
distribution, F555W$-$F814W $\simeq -0.4$ agrees with the color 
predicted for an infinitely hot, unreddened blackbody.}

\figcaption[figure4.ps]
{A comparison of extinction estimates based on PNN colors with those derived
from the nebulae.  The solid line is the relation derived by reddening a
100,000~K Planck curve and folding it through the F555W and F814W filter
response curves.  Both the PNN color-extinctions and the theoretical relation
have been transformed from $E$(F555W$-$F814W) to $E(B$$-$$V)$ using the table
of Holtzman \etal (1995).  PNNs with composite or late-type
spectra have not been plotted.  Note that, although the relation between $c$
and $E(B$$-$$V)$ is good in the mean, there is a substantial dispersion
in the measurements, with $\sigma_{E(B-V)} = 0.16$~mag.}

\figcaption[figure5.ps]
{PNN $V$ magnitudes derived from our {\sl HST\/} measurements compared to the
ground-based $V$ magnitudes of Kaler and collaborators (Shaw \& Kaler
1985, 1989; Jacoby \& Kaler 1989) and Tylenda \etal (1991).  For objects
brighter than $V \simeq 15.4$, the agreement is good, although the scatter is
slightly larger than would be predicted from the individual error bars. At
fainter magnitudes large excursions exist, showing that some ground-based
photometry has been severely contaminated by nebular emission.} 

\figcaption[figure6.ps]
{Positions of the two companion stars to the nucleus of NGC 650-1 in the HR
diagram, under the assumption that the fainter star is on the main sequence. 
Superposed for comparison are the 6, 10, 14, and 18~Gyr solar-metallicity
isochrones from Bertelli \etal (1994).  The error bars reflect photometric
uncertainties only, and do not include the contribution of the uncertain
foreground reddening.  As discussed in the text, since NGC~650-1 likely comes
from a massive progenitor, it seems probable that the PNN and the companions
are not associated.} 

\figcaption[figure7.ps]
{A comparison of directly measured distances to planetary nebulae with those
from four different statistical methods. Distances from resolved binaries are
shown as filled circles, and those from recent PNN parallax measurements are
shown as crosses. The open circles show A~30 and IC~4637, which are only
possible binary associations.  Our upper limit of 0.44~kpc for A~31 from its
resolved companion is shown as a left-facing arrow. All four statistical
methods systematically overestimate the distances to the PNe in our sample,
although for the Cahn, Kaler, \& Stanghellini (1992) method the overestimate
is by only $1 \sigma$. } 

\figcaption[figure8.ps]
{A plot of radio brightness temperature determined from 6~cm observations,
against derived PN radius for three samples of planetary nebulae.  Filled
circles represent PNe with visual binary companions, crosses show PNe with
trigonometric parallax measurements, and open circles show PNe with distances
from Galactic extinction and H~I absorption measurements. The solid line is
the calibration relation for the Van de Steene \& Zijlstra (1995) statistical
distance scale; the dotted line with two segments is that for the statistical
distance scale of Cahn, Kaler, \& Stanghellini (1992).  Note that while PNe
with ISM-based distances obey the statistical relations, the binary and
astrometric PNe have systematically fainter brightness temperatures, and a
much larger amount of scatter.  The data suggest that previous statistical
distance scales have been affected by selection effects in the calibration
dataset, and do not work well for low-surface-brightness nebulae.} 

\newpage


\end{document}


\begin{deluxetable}{lcccc}
\tablenum{1}
\tablewidth{0pt}
\tablecaption{Planetary Nebulae Observed in Cycle 3}
\tablehead{
&\colhead{PN G} &\colhead{Date} 
&\multicolumn{2}{c}{Exposure Times (s)} \\
\colhead{Name} &\colhead{Designation} &\colhead{of Obs.} &\colhead{F555W}
&\colhead{F785LP} }
\startdata
A 24          &$217.1 +14.7$  &1993 Sep 25    &230    &1200 \nl
A 30          &$208.5 +33.2$  &1993 Nov 03    &35     &160  \nl
A 46          &$055.4 +16.0$  &1993 Aug 22    &\dots  &350  \nl
A 63          &$053.8 -03.0$  &1993 Oct 21    &50     &230  \nl
A 82          &$114.0 -04.6$  &1993 Aug 16    &60     &300  \nl
EGB 5         &$211.9 +22.6$  &1993 Nov 07    &\dots  &100  \nl
H 3-75        &$193.6 -09.5$  &1993 Aug 18    &60     &300  \nl
He 1-5        &$060.3 -07.3$  &1993 Nov 16    &\dots  &40   \nl
He 2-36       &$279.6 -03.1$  &1993 Oct 24    &8      &40   \nl
HFG 1         &$136.3 +05.5$  &1993 Sep 24    &\dots  &70   \nl
IC 289        &$138.8 +02.8$  &1993 Oct 10    &\dots  &800  \nl
IC 1747       &$130.2 +01.3$  &1993 Aug 17    &\dots  &400  \nl
IC 2448       &$285.7 -14.9$  &1993 Sep 11    &\dots  &160  \nl
IC 4637       &$345.4 +00.1$  &1993 Sep 22    &\dots  &30   \nl
IC 4997       &$058.3 -10.9$  &1993 Sep 22    &\dots  &180  \nl
K 1-16        &$094.0 +27.4$  &1993 Sep 26    &\dots  &350  \nl
K 2-15        &$263.2 +00.4$  &1993 Oct 24    &2      &10   \nl
LoTr 5        &$339.9 +88.4$  &1993 Nov 25    &0.2    &1    \nl
M 1-2         &$133.1 -08.6$  &1993 Sep 24    &14     &70   \nl
M 1-26        &$358.9 -00.7$  &1993 Sep 20    &\dots  &40   \nl
NGC 1514      &$165.5 -15.2$  &1993 Oct 22    &0.35   &1.8  \nl
NGC 3587      &$148.4 +57.0$  &1993 Oct 24    &\dots  &800  \nl
NGC 6905      &$061.4 -09.5$  &1993 Oct 29    &\dots  &600  \nl
NGC 7009      &$037.7 -34.5$  &1993 Oct 20    &\dots  &40   \nl
NGC 7026      &$089.0 +00.3$  &1993 Oct 24    &\dots  &160  \nl
NGC 7094      &$066.7 -28.2$  &1993 Nov 16    &\dots  &100  \nl
NGC 7293      &$036.1 -57.1$  &1993 Nov 02    &\dots  &70   \nl
PuWe 1        &$158.9 +17.8$  &1993 Oct 24    &80     &400  \nl
Vy 2-2        &$045.4 -02.7$  &1993 Sep 16    &\dots  &230  \nl
\enddata
\end{deluxetable}

\begin{deluxetable}{lllcccc}
\tablenum{2}
\tablewidth{0pt}
\tablecaption{Planetary Nebulae Observed in Cycle 5}
\tablehead{
&\colhead{PN G} &\colhead{Date} 
&\multicolumn{4}{c}{--- Exposure Times (s) ---} \\
\colhead{Name} &\colhead{Designation} &\colhead{of Obs.} &\colhead{F555W-1}
&\colhead{F555W-2} &\colhead{F814W-1} &\colhead{F814W-2} }
\startdata
A 7           &$215.5 -30.8$  &1995 Aug 17   &50     &100    &200    &350     \nl
A 16          &$153.7 +22.8$  &1995 Sep 15   &300    &\dots  &400    &400     \nl
A 21          &$205.1 +14.2$  &1995 Sep 29   &80     &160    &350    &500     \nl
A 28          &$158.8 +37.1$  &1995 Sep 24   &300    &\dots  &400    &400     \nl
A 31          &$219.1 +31.2$  &1995 Nov 19   &50     &100    &200    &350     \nl
A 33          &$238.0 +34.8$  &1996 Feb 13   &50     &100    &200    &350     \nl
A 34          &$248.7 +29.5$  &1995 Nov 27   &100    &\dots  &400    &\dots   \nl
A 39          &$047.0 +42.4$  &1995 Jul 28   &60     &120    &230    &400     \nl
A 43          &$036.0 +17.6$  &1995 Jul 28   &26     &50     &100    &160     \nl
A 61          &$077.6 +14.7$  &1995 Jul 24   &300    &\dots  &400    &400     \nl

A 65          &$017.3 -21.9$  &1995 Aug 12   &80     &160    &300    &500     \nl
A 66          &$019.8 -23.7$  &1995 Aug 13   &300    &\dots  &400    &400     \nl
A 72          &$059.7 -18.7$  &1995 Aug 31   &80     &180    &350    &\dots   \nl
A 74          &$072.7 -17.1$  &1995 Aug 20   &230    &\dots  &400    &400     \nl
A 78          &$081.2 -14.9$  &1995 Oct 25   &6      &12     &23     &40      \nl
EGB 1         &$124.0 +10.7$  &1995 Sep 27   &100    &\dots  &400    &\dots   \nl
He 2-131      &$315.1 -13.0$  &1995 Jul 31   &0.8    &1.6    &3.5    &5       \nl
He 2-138      &$320.1 -09.6$  &1996 Aug 11   &0.8    &1.6    &3      &5       \nl
IC 418        &$215.2 -24.2$  &1995 Sep 24   &0.4    &0.8    &1.6    &2.6     \nl

IC 2149       &$166.1 +10.4$  &1995 Nov 06   &1      &2      &4      &7       \nl
IC 2448       &$285.7 -14.9$  &1995 Oct 10   &16     &30     &60     &100     \nl
IC 3568       &$123.6 +34.5$  &1995 Aug 04   &5      &10     &20     &30      \nl
IC 4406       &$319.6 +15.7$  &1995 Jul 30   &300    &\dots  &400    &400     \nl
IC 4593       &$025.3 +40.8$  &1995 Aug 28   &1      &2      &4      &7       \nl
IC 4637       &$345.4 +00.1$  &1995 Sep 24   &3.5    &7      &14     &23      \nl
IC 4997       &$058.3 -10.9$  &1995 Sep 14   &20     &40     &80     &120     \nl
IC 5148-50    &$002.7 -52.4$  &1995 Sep 02   &50     &100    &200    &350     \nl
IsWe 1        &$149.7 -03.3$  &1995 Nov 08   &140    &\dots  &500    &\dots   \nl
Jn 1          &$104.2 -29.6$  &1995 Jul 29   &100    &180    &350    &\dots   \nl

JnEr 1        &$164.8 +31.1$  &1995 Sep 19   &180    &\dots  &350    &350     \nl
K 1-2         &$253.5 +10.7$  &1995 Sep 26   &100    &\dots  &400    &\dots   \nl
K 1-14        &$045.6 +24.3$  &1996 Jan 22   &120    &\dots  &500    &\dots   \nl
K 1-22        &$283.6 +25.3$  &1995 Aug 08   &80     &180    &350    &\dots   \nl
K 1-27        &$286.8 -29.5$  &1995 Dec 04   &160    &\dots  &600    &\dots   \nl
Lo 4          &$274.3 +09.1$  &1995 Sep 30   &140    &\dots  &\dots  &\dots   \nl
Lo 8          &$310.3 +24.7$  &1995 Sep 09   &5      &10     &20     &35      \nl
M 2-9         &$010.8 +18.0$  &1997 May 24   &12     &23     &100    &180     \nl
Mz 2          &$329.3 -02.8$  &1996 Aug 19   &200    &200    &400    &400     \nl
NGC 40        &$120.0 +09.8$  &1995 Nov 10   &1.4    &3      &6      &10      \nl

NGC 650-1     &$130.9 -10.5$  &1995 Aug 05   &80     &160    &300    &500     \nl
NGC 1360      &$220.3 -53.9$  &1995 Dec 16   &1.2    &2.3    &4      &8       \nl
NGC 1501      &$144.5 +06.5$  &1995 Sep 07   &20     &40     &80     &140     \nl
NGC 1535      &$206.4 -40.5$  &1995 Nov 25   &2.6    &5      &10     &16      \nl
NGC 2022      &$196.6 -10.9$  &1995 Aug 11   &80     &140    &300    &500     \nl
NGC 2346      &$215.6 +03.6$  &1995 Oct 27   &1.2    &2.6    &5      &8       \nl
NGC 2371-2    &$189.1 +19.8$  &1995 Oct 24   &26     &50     &100    &200     \nl
NGC 2392      &$197.8 +17.3$  &1995 Nov 20   &0.5    &1      &2      &3.5     \nl
NGC 2440      &$234.8 +02.4$  &1995 Nov 18   &350    &\dots  &400    &400     \nl
NGC 2452      &$243.3 -01.0$  &1995 Sep 04   &400    &\dots  &400    &400     \nl

NGC 2610      &$239.6 +13.9$  &1995 Oct 10   &70     &140    &260    &400     \nl
NGC 2792      &$265.7 +04.1$  &1995 Aug 11   &260    &\dots  &400    &400     \nl
NGC 2867      &$278.1 -05.9$  &1995 Jul 30   &70     &140    &300    &500     \nl
NGC 3132      &$272.1 +12.3$  &1995 Dec 04   &0.35   &0.7    &1.4    &2.3     \nl
NGC 3195      &$296.6 -20.0$  &1995 Oct 05   &80     &180    &350    &\dots   \nl
NGC 3242      &$261.0 +32.0$  &1996 Feb 15   &2.6    &5      &10     &18      \nl
NGC 3918      &$294.6 +04.7$  &1995 Oct 18   &60     &120    &260    &400     \nl
NGC 4361      &$294.1 +43.6$  &1996 Dec 02   &6      &12     &26     &40      \nl
NGC 5189      &$307.2 -03.4$  &1995 Sep 02   &30     &60     &120    &200     \nl
NGC 5307      &$312.3 +10.5$  &1995 Jul 28   &23     &50     &100    &160     \nl

NGC 5315      &$309.1 -04.3$  &1995 Jul 25   &20     &40     &80     &120     \nl
NGC 5882      &$327.8 +10.0$  &1995 Jul 26   &8      &16     &30     &50      \nl
NGC 5979      &$322.5 -05.2$  &1997 Feb 07   &40     &80     &180    &300     \nl
NGC 6026      &$341.6 +13.7$  &1995 Sep 27   &7      &14     &26     &40      \nl
NGC 6058      &$064.6 +48.2$  &1995 Oct 22   &12     &23     &40     &80      \nl
NGC 6153      &$341.8 +05.4$  &1997 Jun 01   &80     &180    &350    &350     \nl
NGC 6309      &$009.6 +14.8$  &1995 Aug 26   &140    &\dots  &600    &\dots   \nl
NGC 6369      &$002.4 +05.8$  &1995 Aug 09   &80     &160    &300    &500     \nl
NGC 6543      &$096.4 +29.9$  &1995 Nov 22   &1      &2      &4      &6       \nl
NGC 6578      &$010.8 -01.8$  &1995 Aug 16   &70     &140    &300    &500     \nl

NGC 6629      &$009.4 -05.0$  &1995 Aug 16   &5      &10     &20     &35      \nl
NGC 6720      &$063.1 +13.9$  &1995 Sep 17   &60     &120    &260    &400     \nl
NGC 6781      &$041.8 -02.9$  &1995 Jul 24   &180    &\dots  &350    &350     \nl
NGC 6790      &$037.8 -06.3$  &1996 Aug 30   &12     &23     &50     &80      \nl
NGC 6804      &$045.7 -04.5$  &1995 Sep 18   &18     &35     &70     &120     \nl
NGC 6826      &$083.5 +12.7$  &1995 Aug 21   &0.6    &1.2    &2.3    &4       \nl
NGC 6853      &$060.8 -03.6$  &1995 Oct 31   &12     &26     &50     &80      \nl
NGC 6884      &$082.1 +07.0$  &1995 Oct 13   &80     &180    &350    &\dots   \nl
NGC 6891      &$054.1 -12.1$  &1995 Nov 20   &3      &6      &12     &20      \nl
NGC 6894      &$069.4 -02.6$  &1995 Jul 28   &500    &\dots  &400    &400     \nl

NGC 7008      &$093.4 +05.4$  &1995 Aug 24   &6      &12     &26     &40      \nl
NGC 7009      &$037.7 -34.5$  &1997 May 28   &4      &120    &18     &180     \nl
NGC 7027      &$084.9 -03.4$  &1995 Aug 21   &100    &200    &400    &\dots   \nl
NGC 7662      &$106.5 -17.6$  &1995 Sep 27   &6      &12     &26     &40      \nl
PHL 932       &$125.9 -47.0$  &1995 Jul 28   &2.3    &5      &10     &16      \nl
RX J2117+34   &$080.3 -10.4$  &1995 Nov 21   &6      &12     &26     &40      \nl
Sp 3          &$342.5 -14.3$  &1995 Sep 13   &3.5    &7      &14     &23      \nl
Th 2-A        &$306.4 -00.6$  &1995 Sep 22   &20     &40     &80     &140     \nl
\enddata
\end{deluxetable}

\begin{deluxetable}{lcccccccc}
\def\pht{{\phantom{$>$}}}
\footnotesize
\tablenum{3}
\tablewidth{0pt}
\tablecaption{Candidate Companion Stars}
\tablehead{
&\colhead{Separation} &\colhead{J2000 P.A.}
& &\colhead{F555W$-$} &&&\multicolumn{2}{c}{Probability} \\
\colhead{Name} &\colhead{($''$)} &\colhead{(deg)}
&\colhead{F814W} &\colhead{F814W} &\colhead{$V$} 
&\colhead{$V$$-$$I$} &\colhead{Observed} &\colhead{$+ 1\sigma$} }
\startdata
A 7        &0.91 &250 &21.16 &$>$1.20\pht &$>$22.34\pht &$>$1.21\pht &0.005 &0.006 \\
A 24       &3.33 &159 &17.27\tablenotemark{a} &1.34  &18.55 &1.16  &0.021 &0.033 \\
A 30       &5.25 &144 &16.33\tablenotemark{a} &1.11  &17.40 &0.97  &0.035 &0.058 \\
A 31       &0.26 &242 &18.94 &$>$3.12\pht &$>$22.17\pht &$>$3.17\pht &0.001 &0.001 \\
A 33       &1.82 &209 &15.92 &1.16  &17.06 &1.18 &0.017 &0.021 \\
A 63\tablenotemark{a}  &2.82 & 94 &14.61 &1.31  &15.87 &1.13 &0.015 &0.024 \\
IC 4637    &2.42 &330 &13.22 &1.39  &14.60 &1.41 &0.034 &0.041 \\
K 1-14     &0.36 &242 &17.61 &1.03  &18.63 &1.04 &0.001 &0.001 \\
K 1-22     &0.35 &218 &16.02 &1.11  &17.13 &1.12 &0.001 &0.001 \\
K 1-27     &0.56 &315 &20.93 &0.36  &21.29 &0.37 &0.009 &0.010 \\
Mz 2       &0.28 &218 &15.85 &1.23  &17.06 &1.25 &0.000 &0.000 \\
NGC 650-1\tablenotemark{b}  &1.34 &185 &17.42 &1.06  &18.45 &1.07 &0.047 &0.051 \\
                         &1.45 &190 &18.22 &1.06  &19.25 &1.07 &      &      \\
NGC 1535   &1.04 &334 &16.75 &0.93  &17.65 &0.94 &0.002 &0.002 \\
NGC 2392   &2.65 &213 &16.85 &$>$0.51\pht &$>$17.34\pht &$>$0.52\pht &0.012 &0.017 \\
NGC 2610   &0.61 &277 &21.89 &$>$0.78\pht &$>$22.64\pht &$>$0.79\pht &0.008 &0.009 \\
NGC 3132   &1.71 &47  &10.00 &0.12  &10.11 &0.12 &0.001 &0.001 \\
NGC 7008   &0.42 &241 &13.10 &1.32  &14.40 &1.34 &0.001 &0.001 \\
PuWe 1\tablenotemark{b}  &5.20 &119 &13.20\tablenotemark{a} &1.60  &14.74 &1.38  &0.034 &0.057\\
                         &5.45 &113 &15.89 &3.43  &19.22 &2.92 &      &      \\
Sp 3       &0.31 &42  &16.06 &0.82  &16.86 &0.83 &0.002 &0.002 \\

\enddata
\tablenotetext{a}{Magnitude through the WF/PC F785LP filter}
\tablenotetext{b}{The companion to the PNN is itself a binary}
\end{deluxetable}

\begin{deluxetable}{lcccc}
\tablenum{4}
\tablewidth{0pt}
\tablecaption{$V$ Magnitudes, Colors, and Extinctions of PNNs}
\tablehead{
&\colhead{PN G} \\
\colhead{Name} &\colhead{Designation} &\colhead{$V$} &\colhead{$V$$-$$I$} 
&\colhead{$c$} }
\startdata
A 7           &$215.5 -30.8$  &15.58  &$-0.34$  &\dots \nl
A 16          &$153.7 +22.8$  &18.70  &$-0.14$  &0.47  \nl
A 21          &$205.1 +14.2$  &16.05  &$-0.33$  &0.18  \nl
A 24          &$217.1 +14.7$  &17.48  &$-0.38$  &0.48  \nl
A 28          &$158.8 +37.1$  &16.43  &$-0.44$  &0.00  \nl
A 30          &$208.5 +33.2$  &14.38  &$-0.15$  &0.00  \nl
A 31          &$219.1 +31.2$  &15.53  &$-0.42$  &0.00  \nl
A 33          &$238.0 +34.8$  &16.03  &$-0.44$  &0.13  \nl
A 34          &$248.7 +29.5$  &16.47  &$-0.35$  &0.20  \nl
A 39          &$047.0 +42.4$  &15.72  &$-0.24$  &0.00  \nl

A 43          &$036.0 +17.6$  &14.77  &$-0.09$  &0.59  \nl
A 61          &$077.6 +14.7$  &17.47  &$-0.22$  &\dots \nl
A 63          &$053.8 -03.0$  &15.14  &$+0.36$  &0.71  \nl
A 65          &$017.3 -21.9$  &15.80  &$+0.41$  &0.00  \nl
A 66          &$019.8 -23.7$  &18.17  &$-0.12$  &\dots \nl
A 72          &$059.7 -18.7$  &16.10  &$-0.33$  &0.54  \nl
A 74          &$072.7 -17.1$  &17.44  &$-0.03$  &\dots \nl
A 78          &$081.2 -14.9$  &13.26  &$-0.19$  &0.18  \nl
EGB 1         &$124.0 +10.7$  &16.39  &$-0.13$  &\dots \nl
H 3-75        &$193.6 -09.5$  &14.24  &$+1.16$  &\dots \nl

He 2-36       &$279.6 -03.1$  &11.48  &$+0.83$  &1.18  \nl
He 2-131      &$315.1 -13.0$  &10.97  &$+0.08$  &0.20  \nl
He 2-138      &$320.1 -09.6$  &11.03  &$+0.06$  &0.28  \nl
IC 418        &$215.2 -24.2$  &10.23  &$+0.01$  &0.32  \nl
IC 2149       &$166.1 +10.4$  &11.34  &$+0.02$  &0.40  \nl
IC 2448       &$285.7 -14.9$  &14.26  &$-0.28$  &0.11  \nl
IC 3568       &$123.6 +34.5$  &12.97  &$-0.10$  &0.19  \nl
IC 4406       &$319.6 +15.7$  &17.38  &$-0.14$  &0.28  \nl
IC 4593       &$025.3 +40.8$  &11.33  &$-0.19$  &0.07  \nl
IC 4637       &$345.4 +00.1$  &12.70  &$+0.55$  &1.00  \nl

IC 5148-50    &$002.7 -52.4$  &16.16  &$-0.43$  &0.38  \nl
IsWe 1        &$149.7 -03.3$  &16.53  &$-0.13$  &\dots \nl
Jn 1          &$104.2 -29.6$  &16.17  &$-0.29$  &0.08  \nl
JnEr 1        &$164.8 +31.1$  &17.16  &$-0.40$  &0.00  \nl
K 1-2         &$253.5 +10.7$  &16.83  &$+0.20$  &\dots \nl
K 1-14        &$045.6 +24.3$  &16.21  &$-0.27$  &0.00  \nl
K 1-22        &$283.6 +25.3$  &16.83  &$-0.31$  &0.12  \nl
K 1-27        &$286.8 -29.5$  &16.13  &$-0.33$  &0.28  \nl
K 2-15        &$263.2 +00.4$  &12.04  &$+1.28$  &\dots \nl
Lo 4          &$274.3 +09.1$  &16.58  &\dots    &\dots \nl
Lo 8          &$310.3 +24.7$  &13.00  &$-0.33$  &\dots \nl

LoTr 5        &$339.9 +88.4$  & 8.95  &$+1.00$  &0.00  \nl
M 1-2         &$133.1 -08.6$  &13.05  &$+1.10$  &1.11  \nl
M 2-9         &$010.8 +18.0$  &14.45  &$+1.29$  &1.10  \nl
Mz 2          &$329.3 -02.8$  &18.32  &$+0.40$  &1.03  \nl
NGC 40        &$120.0 +09.8$  &11.55  &$+0.27$  &0.80  \nl
NGC 650-1     &$130.9 -10.5$  &17.57  &$-0.16$  &0.19  \nl
NGC 1360      &$220.3 -53.9$  &11.34  &$-0.41$  &0.00  \nl
NGC 1501      &$144.5 +06.5$  &14.36  &$+0.56:$  &1.11  \nl
NGC 1514      &$165.5 -15.2$  & 9.52  &$+0.81$  &0.92  \nl
NGC 1535      &$206.4 -40.5$  &12.11  &$-0.35$  &0.11  \nl

NGC 2022      &$196.6 -10.9$  &15.75  &$-0.39$  &0.46  \nl
NGC 2346      &$215.6 +03.6$  &11.27  &$+0.26$  &0.75  \nl
NGC 2371-2    &$189.1 +19.8$  &14.85  &$-0.31:$  &0.21  \nl
NGC 2392      &$197.8 +17.3$  &10.63  &$-0.14$  &0.19  \nl
NGC 2440      &$234.8 +02.4$  &17.49  &$-0.07$  &0.44  \nl
NGC 2452      &$243.3 -01.0$  &17.46  &$+0.51$  &0.68  \nl
NGC 2610      &$239.6 +13.9$  &15.97  &$-0.35$  &0.10  \nl
NGC 2792      &$265.7 +04.1$  &16.89  &$+0.22$  &0.75  \nl
NGC 2867      &$278.1 -05.9$  &16.03  &$+0.20$  &0.47  \nl
NGC 3132      &$272.1 +12.3$  &15.76  &$-0.24$  &0.23  \nl

NGC 3195      &$296.6 -20.0$  &17.78  &$-0.26$  &0.28  \nl
NGC 3242      &$261.0 +32.0$  &12.32  &$-0.34$  &0.12  \nl
NGC 3918      &$294.6 +04.7$  &15.49  &$+0.01$  &0.33  \nl
NGC 4361      &$294.1 +43.6$  &13.26  &$-0.33$  &0.08  \nl
NGC 5189      &$307.2 -03.4$  &14.53  &$+0.18:$  &0.61  \nl
NGC 5307      &$312.3 +10.5$  &14.74  &$+0.03$  &0.51  \nl
NGC 5882      &$327.8 +10.0$  &13.42  &$+0.07$  &0.40  \nl
NGC 5979      &$322.5 -05.2$  &16.37  &$+0.04$  &0.40  \nl
NGC 6026      &$341.6 +13.7$  &13.33  &$+0.13$  &0.66  \nl
NGC 6058      &$064.6 +48.2$  &13.85  &$-0.36$  &0.04  \nl

NGC 6153      &$341.8 +05.4$  &15.55  &$+0.42$  &1.04  \nl
NGC 6309      &$009.6 +14.8$  &16.31  &\dots    &0.85  \nl
NGC 6369      &$002.4 +05.8$  &15.13  &\dots    &1.94  \nl
NGC 6543      &$096.4 +29.9$  &11.29  &$-0.18$  &0.12  \nl
NGC 6578      &$010.8 -01.8$  &15.68  &$+0.78:$  &1.40  \nl
NGC 6629      &$009.4 -05.0$  &12.87  &$+0.51:$  &0.90  \nl
NGC 6720      &$063.1 +13.9$  &15.78  &$-0.38$  &0.20  \nl
NGC 6781      &$041.8 -02.9$  &16.86  &$+0.41$  &1.12  \nl
NGC 6790      &$037.8 -06.3$  &16.13  &$+0.53$  &0.81  \nl
NGC 6804      &$045.7 -04.5$  &14.17:  &$+0.53:$  &0.87  \nl

NGC 6826      &$083.5 +12.7$  &10.68  &$-0.24$  &0.04  \nl
NGC 6853      &$060.8 -03.6$  &14.09  &$-0.46$  &0.11  \nl
NGC 6884      &$082.1 +07.0$  &16.71  &$+0.49$  &0.81  \nl
NGC 6891      &$054.1 -12.1$  &12.34  &$-0.16$  &0.23  \nl
NGC 6894      &$069.4 -02.6$  &18.32  &$+0.33$  &0.79  \nl
NGC 7008      &$093.4 +05.4$  &13.89  &$+0.21$  &0.66  \nl
NGC 7009      &$037.7 -34.5$  &12.87  &$-0.23$  &0.14  \nl
NGC 7027      &$084.9 -03.4$  &16.53  &$+0.47:$  &1.31  \nl
NGC 7662      &$106.5 -17.6$  &14.00  &$-0.23$  &0.17  \nl
PHL 932       &$125.9 -47.0$  &12.12  &$-0.31$  &\dots \nl

PuWe 1        &$158.9 +17.8$  &15.64  &$-0.27$  &0.23  \nl
RX J2117+34   &$080.3 -10.4$  &13.10  &$-0.40$  &\dots \nl
Sp 3          &$342.5 -14.3$  &13.20  &$-0.19$  &\dots \nl
Th 2-A        &$306.4 -00.6$  &17.08  &$+0.71$  &1.07  \nl

\enddata
\end{deluxetable}

\begin{deluxetable}{lccc}
\tablenum{5}
\tablewidth{0pt}
\tablecaption{$I$ Magnitudes and Extinctions of PNNs}
\tablehead{
&\colhead{PN G} \\
\colhead{Name} &\colhead{Designation} &\colhead{$I$} &\colhead{$c$} }
\startdata
A 46          &$055.4 +16.0$  &15.48 &0.00 \nl
EGB 5         &$211.9 +22.6$  &14.36 &\dots \nl
He 1-5        &$060.3 -07.3$  &10.22 &0.60 \nl
HFG 1         &$136.3 +05.5$  &13.00 &0.55 \nl
IC 289        &$138.8 +02.8$  &15.92 &1.24 \nl
IC 1747       &$130.2 +01.3$  &15.45 &0.93 \nl
IC 4997       &$058.3 -10.9$  &11.91 &0.45 \nl
K 1-16        &$094.0 +27.4$  &15.48 &0.57 \nl
M 1-26        &$358.9 -00.7$  &11.32 &1.48 \nl
NGC 3587      &$148.4 +57.0$  &17.15 &0.07 \nl
NGC 6905      &$061.4 -09.5$  &14.74 &0.19 \nl
NGC 7026      &$089.0 +00.3$  &13.62 &0.76 \nl
NGC 7094      &$066.7 -28.2$  &13.92 &0.31 \nl
NGC 7293      &$036.1 -57.1$  &14.03 &0.00 \nl
Vy 2-2        &$045.4 -02.7$  &12.12 &1.39 \nl

\enddata
\end{deluxetable}

\begin{deluxetable}{ccccc}
\tablenum{6}
\tablewidth{0pt}
\tablecaption{Adopted Main Sequence\tablenotemark{a}}
\tablehead{
\colhead{$V$$-$$I$} &\colhead{$M_V$} & &\colhead{$V$$-$$I$} &\colhead{$M_V$} }
\startdata
$-0.10$\phantom{$-$} &0.0 &\phantom{space}&2.40  &10.53 \nl
0.00  &0.77    &&2.50  &10.85 \nl
0.10  &1.50    &&2.60  &11.16 \nl
0.20  &2.15    &&2.70  &11.55 \nl
0.30  &2.75    &&2.80  &12.10 \nl
0.40  &3.32    &&2.90  &12.70 \nl
0.50  &3.85    &&3.00  &13.22 \nl
0.60  &4.36    &&3.10  &13.65 \nl
0.70  &4.85    &&3.20  &14.02 \nl
0.80  &5.35    &&3.30  &14.35 \nl
0.90  &5.80    &&3.40  &14.64\nl
1.00  &6.18    &&3.50  &14.91 \nl
1.10  &6.50    &&3.60  &15.19 \nl
1.20  &6.81    &&3.70  &15.47 \nl
1.30  &7.12    &&3.80  &15.75 \nl
1.40  &7.42    &&3.90  &16.03 \nl
1.50  &7.73    &&4.00  &16.30 \nl
1.60  &8.04    &&4.10  &16.58 \nl
1.70  &8.35    &&4.20  &16.87 \nl
1.80  &8.66    &&4.30  &17.14 \nl
1.90  &8.97    &&4.40  &17.39 \nl
2.00  &9.28    &&4.50  &17.70 \nl
2.10  &9.58    &&4.60  &18.14 \nl
2.20  &9.88    &&4.70  &18.65 \nl
2.30  &10.20   \nl

\enddata
\tablenotetext{a}{From Harris (1998)}
\end{deluxetable}
\begin{deluxetable}{lcccccccc}
\footnotesize
\tablewidth{0pt}
\def\pht{{\phantom{$>$}}}
\tablenum{7}
\tablewidth{0pt}
\tablecaption{Binary PN Distances and Other Properties}
\tablehead{
& & & &\colhead{Distance}  &\colhead{Separation} 
&\colhead{Nebula}  &\colhead{PNN} &\colhead{PN} \\ 
\colhead{Name} &\colhead{$E(B$$-$$V)$} &\colhead{$(m$$-$$M)_0$}
&\colhead{$\sigma(m-M)$}  &\colhead{(kpc)} &\colhead{(A.U.)}      
&\colhead{Diameter (pc)} &\colhead{$M_V$} &\colhead{$M_{5007}$} } 
\startdata
\multicolumn{9}{l}{Probable Associations} \\
A 31  &0.000 &$<$8.24\pht &\dots &$<$0.44\pht &$<$115\pht &$<$2.09\pht &$>$7.29\pht  &\dots \\
A 33  &0.000 &10.32 &0.31 &1.16 &2110 &1.52 &5.71  &2.82 \\
K 1-14     &0.050  &12.39  &0.33 &3.00 &1080 &0.68  &3.66  &3.47  \\
K 1-22     &0.076  &10.63  &0.33 &1.33 &470  &1.16  &5.96  &3.10  \\
K 1-27     &0.052  & 8.35  &0.53 &0.47 &260  &0.10  &7.61  &8.66  \\
Mz 2       &0.654  &11.67  &0.48 &2.16 &600  &0.24  &4.54  &$-1.79$ \\
NGC 1535   &0.061  &11.81  &0.37 &2.31 &2400 &0.23  &0.10  &$-2.46$ \\
NGC 3132   &0.143  &9.42  &0.71 &0.77 &1310 &0.11  &5.87 &$-0.06$  \\
NGC 7008   &0.457  &7.84  &0.43 &0.37 &160 &0.15  &4.56  &1.82   \\
Sp 3       &0.159  &11.88 &0.53 &2.38 &740 &0.41  &0.80  &1.98   \\
\\
\multicolumn{9}{l}{Possible Associations} \\
A 7 &0.049  &$<$15.52\pht &\dots &$<$12.70\pht &$<$11520\pht &$<$46.8\pht &$>-$0.10\pht &\dots \\
A 30         &0.092  &11.52 &0.38  & 2.02 &10580 &1.24 &2.57  &1.20  \\
A 63         &0.488  &10.43 &0.56  & 1.22 & 3440 &0.24 &3.20  &5.17  \\
IC 4637   &0.701  & 8.50 &0.45 &0.50 &1210  &0.05 &1.94  &0.92 \\
NGC 2392 &0.162 &$<$14.03\pht &\dots &$<$6.41\pht &$<$16970\pht &$<$0.61\pht &$>-$3.93\pht &$>-$5.23\pht \\
NGC 2610 &0.053 &$<$17.54\pht &\dots &$<$32.18\pht &$<$19630\pht &$<$5.93\pht &$>-$1.74\pht &$>-$4.30\pht \\
\\
\multicolumn{9}{l}{Doubtful Associations} \\
A 24         &0.014  &11.88 &0.31  & 2.38 &7920  &4.1  &$5.55$  &2.65  \\
NGC 650-1 &0.155  &13.07 &0.38 &4.12  &5520 &1.34 &3.99  &$-3.34$ \\
         &       &      &     &      &5970 &     &      &        \\
PuWe 1   &0.130 &7.47 &0.31 &0.31 & 1620 &1.82  &7.77  &\dots    \\
         &0.130 &6.89 &0.43 &0.24 & 1300 &1.39  &8.35  &\dots    \\  

\enddata
\end{deluxetable}

\begin{deluxetable}{lccccccc}
\tablenum{8}
\tablewidth{0pt}
\tablecaption{Statistical vs.~Measured Distances}
\tablehead{
&\colhead{Radius}  &\colhead{F(5 GHz) d$\nu$} 
&\multicolumn{4}{c}{--- Statistical Distances (kpc) ---} & Measured\\
\colhead{Name} &\colhead{($''$)} &\colhead{(Jy)} &\colhead{CKS}
&\colhead{VdSZ} &\colhead{Maciel} &\colhead{Zhang} &\colhead{Distance (kpc)} }
\startdata
\noalign{(a) Measured from Resolved Companion}
A 30  &63.5   &0.0023  &1.69  &4.97  &4.37  &10.38 &2.02  \\
A 31  &486.0  &0.1019  &0.23  &0.72  &0.63  &1.53  &0.44  \\
A 33  &134.0  &0.0140  &0.75  &2.11  &1.86  &4.32  &1.16  \\
A 63  &\dots  &\dots  &\dots  &\dots &\dots &\dots &1.22  \\
IC 4637   &9.3    &0.1325  &2.37  &2.14  &2.00  &2.59  &0.50  \\
K 1-14    &23.5   &0.0014  &3.38  &7.97  &7.15  &15.00 &3.00  \\
K 1-22    &90.5   &0.0115  &0.99  &2.55  &2.26  &4.99  &1.33  \\
K 1-27    &\dots  &\dots   &\dots &\dots &\dots &\dots &0.47  \\
Mz 2      &11.4   &0.0750  &2.35  &2.46  &2.31  &3.18  &2.16  \\
NGC 246   &112.0  &0.2480  &0.47  &0.81  &0.75  &1.33  &0.49  \\
NGC 1535  &9.2    &0.160   &2.30  &2.01  &1.87  &2.39  &2.31  \\
NGC 3132  &22.5   &0.230   &1.25  &1.35  &1.27  &1.78  &0.77  \\
NGC 7008  &42.8   &0.217   &0.86  &1.14  &1.06  &1.65  &0.37  \\
Sp 3      &17.8   &0.0610  &1.88  &2.31  &2.16  &3.23  &2.38  \\
\noalign{(b) Measured from Trigonometric Parallax}
A 21  &307.5  &0.3270  &0.24  &0.55  &0.49  &1.01  &0.54  \\
A 35  &386.0  &0.255   &0.22  &0.56  &0.50  &1.08  &0.13  \\
NGC 1514  &80.0   &0.288   &0.56  &0.86  &0.79  &1.32  &0.18  \\
NGC 6853  &165.0  &1.3249  &0.27  &0.40  &0.37  &0.62  &0.38  \\
NGC 7293  &300.0  &1.2919  &0.19  &0.34  &0.31  &0.57  &0.21  \\
PHL 932   &135.0  &0.0100  &0.80  &2.37  &2.08  &4.97  &0.11  \\
PuWe 1    &1200.0 &0.0847  &0.14  &0.58  &0.49  &1.43  &0.43  \\
\enddata
\end{deluxetable}